\documentclass[a4paper,12pt,reqno]{amsart}
\usepackage{amssymb, amscd,enumerate,enumitem,color, hyperref, tikz, mathtools, relsize, bm, graphicx}               
\usepackage{color}
\usepackage[normalem]{ulem}

\usepackage[british]{babel}
\usepackage[mathscr]{euscript}

\usepackage[makeroom]{cancel}

\makeatletter
\newdimen\rh@wd
\newdimen\rh@hta
\newdimen\rh@htb
\newbox\rh@box
\def\rh@measure#1{\setbox\rh@box=\hbox{$#1$}\rh@wd=\wd\rh@box \rh@hta=\ht\rh@box}

\def\widecheck#1{\rh@measure{#1}%
  \setbox\rh@box=\hbox{$\widehat{\vrule height \rh@hta width\z@ \kern\rh@wd}$}%
  \rh@htb=\ht\rh@box \advance\rh@htb\rh@hta \advance\rh@htb\p@
  \ooalign{$\vrule height \ht\rh@box width\z@ #1$\cr
           \raise\rh@htb\hbox{\scalebox{1}[-1]{\box\rh@box}}\cr}}
\makeatother

\makeatletter
\@namedef{subjclassname@2020}{%
  \textup{2020} Mathematics Subject Classification}
\makeatother

\makeatletter
\def\subsection{\@startsection{subsection}{3}%
  \z@{.5\linespacing\@plus.7\linespacing}{.1\linespacing}%
  {\normalfont\bfseries}}
\makeatother

\font\smallsmc = cmcsc9
\font\smalltt = cmtt8
\font\smallit = cmti8

\numberwithin{equation}{section}

\theoremstyle{plain}
\newtheorem{theo}{Theorem}[section]
\newtheorem{lem}[theo]{Lemma}
\newtheorem{prop}[theo]{Proposition}

\newtheorem{cor}[theo]{Corollary}

\theoremstyle{definition}
\newtheorem{rem}[theo]{Remark}

\newtheorem{remark}[theo]{Remark}

\newtheorem{definition}[theo]{Definition}

\newenvironment{pf}{\noindent{\it Proof. }}{$\square$\par\medskip}

\theoremstyle{plain}

\theoremstyle{definition}

\newcommand{\beq}{\begin{equation}}
\newcommand{\eeq}{\end{equation}}
\renewcommand{\a}{\alpha}
\renewcommand{\b}{\beta}

\renewcommand{\d}{\delta}

\newcommand{\f}{\varphi}
\newcommand{\g}{\gamma}

\newcommand{\q}{\vartheta}
\renewcommand{\r}{\rho}
\newcommand{\s}{\sigma}
\renewcommand{\t}{\tau}
\newcommand{\x}{\xi}

\newcommand{\D}{\Delta}

\newcommand{\F}{\Phi}
\newcommand{\G}{\Gamma}

\newcommand{\bC}{\mathbb{C}}

\newcommand{\bR}{\mathbb{R}}
\newcommand{\bZ}{\mathbb{Z}}

\newcommand{\bF}{\mathbb{F}}
\newcommand{\bG}{\mathbb{G}}

\newcommand{\ga}{\mathfrak{a}}
\newcommand{\gb}{\mathfrak{b}}
\newcommand{\gc}{\mathfrak{c}}
\newcommand{\gd}{\mathfrak{d}}
\newcommand{\gf}{\mathfrak{f}}
\renewcommand{\gg}{\mathfrak{g}}
\newcommand{\gh}{\mathfrak{h}}

\newcommand{\gp}{\mathfrak{p}}
\newcommand{\gq}{\mathfrak{q}}

\newcommand{\gF}{\mathfrak{F}}

\newcommand\SO{\mathrm{SO}}

\newcommand\Spin{\mathrm{Spin}}
\newcommand\Sp{\mathrm{Sp}}

\newcommand{\cA}{\mathscr{A}}
\newcommand{\cB}{\mathscr{B}}
\newcommand{\cC}{\mathcal{C}}
\newcommand{\cD}{\mathscr{D}}

\newcommand{\cH}{\mathscr{H}}
\newcommand{\ch}{\mathscr{H}}

\newcommand{\cK}{\mathscr{K}}
\newcommand{\cL}{\mathscr{L}}
\newcommand{\cM}{\mathscr{M}}

\newcommand{\cN}{\mathscr{N}}

\newcommand{\cS}{\mathscr{S}}
\newcommand{\cT}{\mathscr{T}}
\newcommand{\cU}{\mathscr{U}}
\newcommand{\cV}{\mathscr{V}}

\newcommand{\cW}{\mathscr{W}}

\newcommand{\p}{\partial}

\renewcommand{\square}{\kern1pt\vbox
{\hrule height 0.6pt\hbox{\vrule width 0.6pt\hskip 3pt
\vbox{\vskip 6pt}\hskip 3pt\vrule width 0.6pt}\hrule height0.6pt}\kern1pt}

\DeclareMathOperator\tr{tr\;}
\DeclareMathOperator\Tr{tr\;}

\DeclareMathOperator\Id{Id}

\DeclareMathOperator{\Span}{span}

\newcommand\Hom{\operatorname{Hom}}
\renewcommand\Re{\operatorname{Re}}
\renewcommand\Im{\operatorname{Im}}
\renewcommand\={:=}

\newcommand{\wt}{\widetilde}

\newcommand{\bt}{\begin{theo}\ \ }
\newcommand{\et}{\end{theo}}
\newcommand{\bp}{\begin{prop}\ \ }
\newcommand{\ep}{\end{prop}}
\newcommand{\bc}{\begin{cor}\ \ }
\newcommand{\ec}{\end{cor}}
\newcommand{\bl}{\begin{lem}\ \ }
\newcommand{\el}{\end{lem}}
\newcommand{\bd}{\begin{definition}}
\newcommand{\ed}{\end{definition}}
\newcommand{\n}{\nabla}

\newcommand{\be}{\begin{equation}}
\newcommand{\ee}{\end{equation}}

\def\<#1,#2>{\langle\,#1,\,#2\,\rangle}

\newcommand{\arr}{\begin{array}{rlll}}
\newcommand{\ea}{\end{array}}
\newcommand{\bea}{\begin{eqnarray}}
\newcommand{\eea}{\end{eqnarray}}
\newcommand{\bean}{\begin{eqnarray*}}
\newcommand{\eean}{\end{eqnarray*}}

\newcommand{\Cl}{{\cC \ell}}
\renewcommand{\Cl}{{\cC \ell}}
\newcommand{\bdot}{{\boldsymbol{\cdot}}}

\def\sideremark#1{\ifvmode\leavevmode\fi\vadjust{
\vbox to0pt{\hbox to 0pt{\hskip\hsize\hskip1em
\vbox{\hsize3cm\tiny\raggedright\pretolerance10000
\noindent #1\hfill}\hss}\vbox to8pt{\vfil}\vss}}}
\begin{document}

\title[On the entropy of  the  BPS black holes]
{Special Vinberg cones\\
 and the entropy of  BPS extremal black holes
 }
\author[D. V. Alekseevsky]{Dmitri Vladimirovich Alekseevsky}
\author[A. Marrani]{Alessio Marrani}
\author[A.  Spiro]{Andrea  Spiro}

\begin{abstract}  We consider the static, spherically symmetric and asymptotically flat    BPS extremal black holes in
 ungauged  $N=2$  $D=4$ supergravity theories, in which the scalar manifold of the vector multiplets is homogeneous.   By  a result of Shmakova on the  BPS   attractor equations,  the entropy of this kind of black holes can be  expressed only in terms of  their electric and magnetic charges, provided that the inverse of  a certain  quadratic map (uniquely determined by  the prepotential of the theory) is given.  This   inverse  was previously  known just for the  cases in which  the  scalar manifold of the theory  is  a homogeneous symmetric space. In this paper
we use Vinberg's theory  of homogeneous cones to  determine an   explicit expression for such  an inverse, under the assumption that  the  scalar manifold is homogeneous, but not necessarily symmetric.  As  immediate consequence,  we get  a formula for the  entropy of   BPS black holes that holds in any model of $N = 2$  supergravity with homogeneous scalar manifold.
\end{abstract}

\maketitle
\setcounter{section}0
\section{Introduction}
The first  purpose of this paper  is to determine an explicit formula which gives  the  Bekenstein-Hawking  entropy of a  static, spherically symmetric and asymptotically flat    BPS extremal black hole in terms of its electric and magnetic charges in
 ungauged  $N=2$  $D=4$ supergravity theory,  under the assumption that the scalar manifold of the vector multiplets is homogeneous. As a secondary goal, we  want  to  offer   a gentle    introduction  to  Vinberg's  theory of homogeneous cones  associated with irreducible invariant cubic polynomials and to  illustrate how  this purely   mathematical theory  can be combined  with some fundamental theoretical
 physics results, such as Bekenstein-Hawking entropy-area formula or  Ferrara, Kallosh and Strominger's  BPS algebraic attractor equations,   to establish new non-trivial  results on  black holes.\par
 \medskip
The paper starts with  a discussion of   the invariant real cubic polynomials  $d(y) = d_{abc} y^a y^b y^c$,  $(y^a) \in \bR^n$,  which are  associated to the  holomorphic prepotentials
\beq\label{cubpol}  F(X) =\frac{d_{abc} X^a X^b X^c}{X^0}\ ,\qquad (X^I) = (X^0, X^a) \in \bC^{n+1}\ ,\eeq that determine  homogeneous scalar manifolds  of  the  vector multiplets of  ungauged  $N=2$  $D=4$ supergravity.  By  known results   on   prepotentials and associated  scalar manifolds  (\cite{WP1, Ce, Co, AC1}), any irreducible cubic polynomial of this kind corresponds to  a rank $3$  homogeneous convex cone $\cV \subset \bR^n$ of dimension $n$. This  is in turn representable as the cone of positive Hermitian matrices   in an appropriate  space of $3\times 3$ matrices, with vector and spinor valued entries,   given by  a special  Vinberg $T$-algebra (see \cite{Vi, ACM} and  \S\ 2.2 - 2.3). Aiming to
 a presentation  that might be  accessible  to any  reader with  no  previous knowledge of   Vinberg's theory, the first section  starts with  a self-contained
exposition of the main definitions and properties of   special Vinberg $T$-algebras, of  the  corresponding   cones  of  positive Hermitian matrices and of  the  invariant cubic polynomials which are important objects associated with  these cones. We also introduce the notions of   dual cones and   associated  dual invariant  homogeneous rational functions of degree $3$. These new objects  are later used  to  determine the explicit general expressions for the inverses to  the quadratic maps, that appear in the explicit entropy formula   for BPS extremal black holes established in the third part.  \par
\smallskip
The second and third parts of the paper provide a short review  of   special K\"ahler geometry and of the  BPS extremal black holes in ungauged Maxwell-Einstein $N = 2$ supergravity with prepotentials of the form \eqref{cubpol}.  The presentation  is  structured  for readers who are not familiar with supergravity.
Here is a short outline.
 Consider  the  class of  metrics on the $4$-dimensional space-time $M$  of the form \cite{Pap, Maj},
 \beq \label{2}
ds^{2}=-e^{2U(r)}dt^2+e^{-2U(r)}\big(dr^2+r^2( d\theta^2+\sin^2 \theta d\varphi^2) \big) .
 \eeq
They are  solutions to  the Euler-Lagrange equations  of the bosonic sector of
 the aforementioned  supergravity
and describe  static, spherically symmetric,
asymptotically flat, dyonic extremal black holes with unique event horizon at $r= 0$ (as implied by extremality).
The magnetic and electric charges of  any such  black hole
are the fluxes of the  electromagnetic  fields  $\bF^I = \bF^I_{\mu \nu} d x^\mu \wedge d x^\nu$ and their duals $ \bG_J = \star  \frac{1}{2} \frac{\delta \cL}{\delta \bF^{J}}$ with respect to  the Lagrangian $\cL$ of the theory,
\beq
p^I \=\frac{1}{4\pi }\int_{S_{\infty }^{2}}\bF^I\ ,\qquad q_J\=\frac{1}{4\pi }\int_{S_{\infty }^{2}} \bG_J\ .\label
{charges-1}
\eeq
Due to its rotational invariance and   time independence, the dynamics of  the scalar and electromagnetic fields of  this kind of black  holes can be described by  means of a reduced $1$-dimensional  Lagrangian,
characterised by  the  {\it effective black hole potential} \cite{FGK}
\beq
V_{BH}(p^I,q_J, z,\overline{z}) \=\left\vert Z\right\vert ^{2}+g^{a\overline{b}}  Z_a \overline{Z_b}\qquad \text{with} \ \ Z_a \= \frac{\p Z}{\p z^a} + \frac{1}{2} \frac{\p \cK}{\p z^a} Z \ , \label{V-N=2}
\eeq
where (i)
  $z = (z^a)$ is the map that   represents the scalar fields and takes values in the scalar manifold $\cS \subset \bC^n$;
(ii)  $g = (g_{a\bar b})$ is the K\"ahler metric of the scalar manifold $\cS$; (iii) $\cK$ is  the K\"ahler potential of $g$,  (iv) $Z = Z(p^I, q_J, z, \overline{z})$  is the $N= 2$ central charge, which is a function    of the magnetic and electric  charges $(p^I, q_J)$  and the scalar fields $z = (z^a)$.  We recall that, for any  extremal black hole \eqref{2},  the central charge function  $Z = Z(p^I, q_J, z, \overline{z})$ satisfies the  identity
\begin{equation}
\left(\begin{matrix} p^I \\q_J \end{matrix} \right)=-2 e^{\frac{\cK}{2}} \text{Im}\left( \overline{Z}\left(\begin{matrix} X^I \\F_J \end{matrix} \right) +g^{a\overline{b}
}Z_a \overline{\left( \frac{\p }{\p  z^b} + \frac{\p \cK}{\p  z^b} \right) \left(\begin{matrix} X^I \\F_J \end{matrix} \right) }\right)\  ,\label{SKG-id}
\end{equation}
where $(X^I) = \begin{pmatrix} X^0 \\ X^a\end{pmatrix} = X^0 \begin{pmatrix} 1 \\ z^a\end{pmatrix} \in \bC^{n+1}$ with $z^a = \frac{X^a}{X^0}$,   and   $F_J \= \frac{\p F}{\p X_J}\big|_{(X^I)}$. \par
For  given  charges $p^I$, $q_J$ in an appropriate set, it is  known that
the  values of the scalar fields $z$ at   $r = 0$    are stable critical points for  $V_{BH}$.
By the famous {\it Attractor
Mechanism} \cite{FKS, St, FK, FK1, FGK},  such  criticality condition  determines   a (locally invertible) relation  between the horizon values  of the scalar fields and   the magnetic and electric charges of the black hole.
\par
Let us now focus on  the  black holes \eqref{2} which are in addition BPS. For them  the following two  facts hold:  (a) the  central charge  $Z_o = Z|_{r = 0}$ at the horizon  verifies the relation $ |Z_o|^2= \frac{A_{H}}{4\pi}$,  where  $ A_H$ denotes the area of the event  horizon surface; (b)  all covariant derivatives $Z_a \= \frac{\p Z}{\p z^a} + \frac{1}{2} \frac{\p \cK}{\p z^a} Z$   are identically zero  at  $r = 0$   and  the effective potential  at the horizon reduces  to
$V_{BH}|_{ r = 0} = |Z_o|^{2}$.  In particular, for such black holes
 the  identities (\ref{SKG-id})  imply  the following
{\it purely algebraic} relations between $Z_o$, the magnetic and electric charges and the values $X_o = (X_o^I) = X_o^0 \begin{pmatrix} 1 \\ z_o^a\end{pmatrix}$,    associated with the  scalar fields $z_o \= z|_{r = 0}$:
\begin{equation} \label{algebraic}
\left(\begin{matrix} p^I \\q_J \end{matrix} \right)=-2 e^{\frac{\cK(z_o)}{2}} \text{Im}\left(\overline{Z_o}\left(\begin{matrix} X^I_o \\F_{o J} \end{matrix} \right) \right) .
\end{equation}
This is  a system of equations that relates  the set of the  $2n+2$ real numbers $p^I$, $q_J $ and the set of the   $n+1$ complex numbers $(Z_o, z^1_o, \ldots, z^n_o)$.
Such a system is always locally solvable in terms of  the second set  \cite{Sh}.
\par
On the other hand,  by the Bekenstein-Hawking entropy-area formula  \cite{Be, Ha},  the above property (a) of the BPS metrics \eqref{2} implies that the entropy of any such black hole is related with the central charge $Z_o$ at the horizon by  \cite{FGK}
\begin{equation}
S  =\frac{A_H}{4}=\pi  |Z_o|^{2}\ .
\label{entr}
\end{equation}
Hence any  (local) inversion of the algebraic relation \eqref{algebraic}  (thus giving $Z_o$ as a function of  $p^I$ and $q_J$)  provides  a formula  for the black hole entropy $S$  in terms of its magnetic and electric charges. The fourth part of our paper is devoted  to the solution of such inversion problem.
\par
\smallskip
More precisely,  in our fourth   section,   we analyse in  detail the correspondence $(z^a_o, Z_o) \mapsto (p^I, q_J)$ determined by  \eqref{algebraic},  which we name {\it BPS map}.  After a brief discussion about  the invertibility of such a map  in the (simpler)  case with  $p^0 = 0$, we tackle the  situations  with  $p^0 \neq 0$. For them, we show that   the  BPS  map is  always a  local diffeomorphism.   This is obtained  using    Shmakova's formulas in \cite{Sh} (which we newly derive in detail by means of  a different  line of arguments) that  reduce the inversion problem for  the BPS map  to the (somehow simpler)  inversion problem  of  the  quadratic map
$$h_d: \bR^n \longrightarrow \bR^{n'} = \Hom(\bR^n, \bR) \ ,\qquad h_d(y) \=  \left(d_{abc} y^b y^c\right)\ .$$
From   these formulas, the BPS map is proved to be a local diffeomorphism by  checking  that  $h_d$ has  a non-vanishing Jacobian   and thus, due to the Inverse Function Theorem, it  is    locally invertible with smooth inverse.  In the cases in which $d(y)$  is an irreducible polynomial  and  is associated  with  homogeneous scalar manifolds,   a   {\it globally defined} inverse  for    $h_d: \bR^n \to \bR^{n*}$  has been  explicitly determined in the first part of this paper.   Therefore, the results of our first part  immediately yield to explicit formulas for    {\it global} inverses to the BPS map, one per each of the  two connected regions corresponding to the condition   $p^0 \neq 0$. These two inverse maps give  expressions for   the horizon values  $Z_o$,   $z_o = (z^a_o)$  in terms of the electric and magnetic charges only. By \eqref{entr}, they also  give  an expression   for the entropy of  the black hole.  \par
\medskip
At the best of our knowledge,  explicit expressions for the inverse map $h_d^{-1}$ and,  consequently,  for the entropy  $S$ of the above BPS black holes,  was so far known only when  $d(y) = d_{abc} y^a y^b y^c$  and the corresponding  prepotential  determine   a {\it homogeneous symmetric} scalar manifold $\cS$.  Our results   can be thus considered as a completion of such  previous   results,    providing  a solution   to the above BPS black hole entropy problem for all cases in which  $\cS$ is homogeneous,  regardless whether it is symmetric or non-symmetric.\par
\smallskip
 The  expression for the entropy  $S$ we obtain for the considered cases is
\beq  \label{quartic}
\begin{split}
&S   =  \pi \sqrt{  I_4}\ ,\  \text{where} \ I_4  = I_4(p^0, p^a, q_0, q_b)\ \text{is   the}\\
& \hskip 4 cm \text{homogeneous rational function of degree $4$}\\[15pt]
& I_4(p^0, p^a, q_0, q_b)
= - \frac{1}{(p^0)^2}  \bigg(( q_0 p^0  +  \langle  q, p \rangle)p^0
-  2 d(p)
\bigg)^2 + \\
&   + \frac{4 }{ (p^0)^2}  \bigg(   (1 + \ga(p^0,p^a,q_0,q_a))
\sqrt{d' \left( h( p ) - \frac{1}{3}p^0q\right) }  + \gb(p^0,p^a,q_0,q_a)\bigg)^2  
\end{split}
\eeq
Here $d'(w) = d^{abc} w_a w_b w_c$ where $d^{abc} = \d^{ae} \d^{bf} \d^{cg} d_{efg}$ and $\ga(p^0,p^a$,$q_0,q_a)$, $\gb(p^0,$ $p^a$,$q_0,q_a)$ are  homogeneous rational functions of degree $0$ and $3$, respectively  (vanishing identically if the scalar manifold is a homogeneous symmetric Hermitian space),  explicitly given in \eqref{534}, and all of these terms are uniquely determined by the invariant cubic polynomial  $d(y)$ and  the  {\it dual  invariant  degree $3$ rational function $d^* = d + \gd$ associated with $d(y)$}, which we define  in Definition \ref{invariant-cubic-polynomial}
and  we  explicitly determine in Theorem \ref{lemma17}.  As we have already pointed out, formula \eqref{quartic} was   previously known  just in the cases in which    $(\cS = G/K, g)$  is a homogeneous symmetric Hermitian space. In  all such  cases it was  observed by the second author  in \cite{Mar}  that, up to a factor, 
the natural extension of the  polynomial $I_4$ to $\bC^{2n+2}$  coincides with the unique   generator of the ring of the  relative  invariants of  the standard  representation of $\bC^* \times \r(G)^\bC = (\bR_+ \times \r(G))^\bC$ on $\bC^{2n+2}$, where   $\r: G \hookrightarrow \Sp_{2n+2}(\bR)$   is an appropriate
embedding of $G$ into  $ \Sp_{2n+2}(\bR)$. This remarkable property is   a  consequence of  the following three facts: (a)   the equations of motion of the  supergravity  that   we  consider are invariant under  the  group $G_{\text{e.m.}}$ of the  {\it electric-magnetic dualities} (also called {\it  generalised duality transformations} or  {\it   $U$-dualities} in supergravity literature);  (b)  $G_{\text{e.m.}}$   is a subgroup of $\Sp_{2n+2}(\bR)$ acting in a standard way on the space $\bR^{2n+2}$ of the magnetic and electric charges of the above  BPS black holes and  admitting a natural  isomorphism  $\s = \r^{-1} : G_{\text{e.m.}} \to G = \operatorname{Iso}^o(\cS, g)  $ with  the identity  component $G = \operatorname{Iso}^o(\cS, g)  $ of the  isometry group  $ \operatorname{Iso}(\cS, g) $; (c) the standard representation of  $\bC^* \times (G_{\text{e.m.}})^\bC $ on  $\bC^{2n +2}$ has an open orbit. Note that (c) is  precisely  the property  that implies   that $I_4$ is the {\it unique}  generator (up to a scaling)    for the ring of the  relative invariants.  \par
Since  (a) and (b)  are true   whenever  $(\cS = G/K, g)$ is a homogeneous (not necessarily  symmetric) manifold,  we immediately have that {\it in all these cases}  $I_4$ is  a relative invariant for the    action    of  $ (\bR_+ \times \r(G))^\bC $,  $G^\bC \subset \Sp_{2n+2}(\bC)$, on $\bC^{2n+2}$.   But  we claim that also   (c)  is true  for any homogeneous scalar manifold. In fact,  from  the explicit expression of the BPS map and recalling the explicit form of the  linear action of $G$ on $\bC^{n+1}$ (which projects onto the scalar manifold $\cS$ under projectivisation), one can directly check  that the  isotropy subgroup of the representation    $\r(G) = G_{\text{e.m.}}$  on $\bR^{2n+2}$  has  one dimension  less than the isotropy $H$ of the scalar manifold $\cS = G/H$. Thus the regular orbits  $\r(G){\cdot} (p^I,q_J) \subset \bR^{2n+2}$   have dimension  $\dim_\bR \r(G)\cdot (p,q) = \dim_\bR \cS + 1 = 2 n +1$, a property which implies (c).  Combining  these three simple observations, we may conclude that  (up to a rescaling)  {\it $I_4$  is the unique generator of the relative invariants of  $ (\bR_+ \times \r(G))^\bC $ for all cases in which $\cS$ is     homogeneous}.   Details on   this and other aspects of the  homogeneous rational function $I_4$ of degree $4$ are left  to a future work.
\par
\noindent{\it Acknowledgements}. The authors are sincerely grateful to the Referee for her/his constructive criticism, who helped to consistently improve the presentation.
\medskip
\section{Geometry of special Vinberg cones and their  duals}
\subsection{Special $T$-algebras and their standard matrix representations}
\subsubsection{Special $T$-algebras}
Let $(V, g_V)$ be a Euclidean vector space with associated Clifford algebra $\Cl(V)$, constructed according to   the  Clifford relation
  $ v{\cdot} w + w \cdot v = - 2 g_V(v, w)$.  Let also  $S = S_0 + S_1$  be  a $\bZ_2$-graded $\Cl(V)$ module
equipped with
 a Euclidean scalar product $g_S$  satisfying the  following two conditions:
 \begin{itemize}
\item[(1)] $g_S(S_0, S_1) = 0$;
\item[(2)] each   Clifford multiplication $ \mu(v, \cdot): S \to S$,   $v \in V$,  is  $g_S$-skew symmetric.
\end{itemize}
In the terminology of \cite{AC, AC1},  a scalar product satisfying these  conditions is called
  {\it symmetric admissible scalar product of type $\t = -1$.}
  \par
We call the space $(S, g_S)$ a {\it  metric  $\Cl(V)$-module}.   \par
\medskip
In the following, to  simplify notation,   we   will denote both  the   metric $g_V + g_S$  of $V+S$ and  the corresponding induced metric on  $V^{\prime }+S^{\prime }=$Hom$\left( V+S,\mathbb{R}\right) $  by  $\langle \cdot, \cdot\rangle $.
We will also denote by
 $(\cdot)^\flat: V + S \to V' + S'$  the  associated isomorphism
$$ v  + s \mapsto v^\flat + s^\flat = \langle v + s, \cdot\rangle  \ ,$$
 by $(\cdot)^\sharp: V' + S' \to V+ S$ the inverse map of $(\cdot)^\flat$ and by
 $\langle \cdot, \cdot \rangle_{V' +S'} = g_{V'} + g_{S'}$ the  scalar product on $V' + S'$,  induced by $(\cdot)^\flat$.  A similar notation will be later used for any other Euclidean vector space.  \par
 \medskip
 \begin{definition}
 The {\it special $T$-algebra determined by $(V, g_V)$ and $(S = S_0 + S_1, g_S)$} is the  direct sum of vector spaces
 \beq \label{Talg} \cA =   (S_0' +V' +  S_1')  + \bR  + \bR + \bR + (V +  S_1 + S_0) \  ,\eeq
  equipped with   the product  ``$\bdot$''  and the Euclidean scalar product $(\cdot, \cdot )$ defined as follows. Consider the notation
\begin{align*}
 & &&\cA_{11} = \cA_{22} = \cA_{33} \= \bR\ ,&&\\
 & \cA_{12} \=  S_0\ ,&&  \cA_{23} \=  V\ ,&&  \cA_{13} \= S_1\ ,\\
 &\cA_{21} \= \cA_{12}'  = S_0'\ , && \cA_{32} \= \cA'_{23} = V'\ ,&&  \cA_{31} \= \cA_{13}' = S_1'
 \end{align*}
and define ``$\bdot$'' as the product such that the only non-trivial multiplications are given by the bilinear maps
  $\bdot : \cA_{ij} \times \cA_{j\ell} \longrightarrow \cA_{i \ell}$
that are determined by the Clifford multiplication $\mu: V \times S \to S$
  as follows:
 \begin{align}
 \label{1.2} & \bdot: \cA_{ii} \times \cA_{i\ell} \to \cA_{i\ell}\ , && \r \bdot a = \r a \ \  (\text{mult. by the real number}\ \r)\ ,\\
\label{1.3} & \bdot: \cA_{ij} \times \cA_{ji} \to \cA_{ii}\ ,  && x \bdot y^\flat  = y \bdot x^\flat \=  \langle x, y\rangle  \ \ (\text{scalar product})\ ,\\
\label{1.4} & \bdot: \cA_{12} \times \cA_{23}  \to \cA_{13} \ , && s_0 \bdot v\= \mu(v, s_0) \ , \\
  \label{1.6} & \bdot: \cA_{31} \times \cA_{12}  \to \cA_{32}\ ,&& \langle s_0^\flat\bdot s_1, v\rangle\= \langle \mu(v, s_0), s_1\rangle \\
\label{1.5} & \bdot: \cA_{23} \times \cA_{31} \to \cA_{21} \ ,&& \langle v \bdot s_1^\flat, s_0\rangle\=  \langle \mu(v,  s_0),    s_1\rangle   \ ,
 \end{align}
 together with the rules
 \beq   \label{1.7}
 \begin{split}
 & \bdot: \cA_{ij} \times \cA_{j j}  \to \cA_{i j}\ ,\qquad a_{ij} \bdot \rho\= \rho \bdot a_{ij}\ ,\\
 &\bdot: \cA_{ij} \times \cA_{j \ell}  \to \cA_{i \ell} \ ,\qquad a_{ij} \bdot a_{j \ell} \= (a_{j\ell}^\ast \bdot a_{ij}^\ast)^\ast\ ,\qquad i < j < \ell
 \end{split}
 \eeq
 where  $(\cdot)^\ast: \cA \to \cA$ is the involutive  linear map such that $\cA_{ij}^* = \cA_{ji}$ with
 $$(\cdot)^\ast|_{\cA_{ij}} = \left\{\begin{array}{ccc} \Id_{\cA_{ii}} & \text{if} & i = j\ ,\\[5pt]
  (\cdot)^\flat & \text{if} & i < j\ ,\\[5pt]
(\cdot)^\sharp & \text{if} &  i >  j\ .
 \end{array}\right.
 $$
 This completely determines all
 of the  remaining  products.\par
 Finally, we set $( \cdot, \cdot )$ to be the Euclidean scalar product on $\cA$, with respect to which all subspaces $\cA_{ij}$ are orthogonal one to the other,  is the standard scalar product  of $\bR$  on each subspace   $\cA_{11} = \cA_{22} = \cA_{33} = \bR$, and  is equal to  $ \langle \cdot, \cdot\rangle_{V' + S'} + \langle \cdot, \cdot\rangle $ on  the subspace $ (S_0' +V' +  S_1')  + (V +  S_1 + S_0)$.
  \end{definition}
  In \cite{AC1}  it is proven that any special $T$-algebra, as defined above,  satisfies the axioms of  $T$-algebras of a rank $3$ in the sense of Vinberg.
  This can be also directly checked, by observing that the  Lie algebra
    $$\mathfrak N = \cA_{12} + \cA_{13} + \cA_{23} = V + S_0 + S_1\ ,$$
  equipped with the Euclidean scalar product $(\cdot, \cdot)_{\cN \times \cN}$,   is an  {\it $N$-algebra of rank 3} in  the sense of \cite[\S III.7]{Vi}.
  Indeed, by Vinberg's results,
any  $T$-algebra of  rank $3$ is uniquely determined by its nilpotent part $(\mathfrak N = \cA_{12} + \cA_{23} + \cA_{13}, (\cdot, \cdot))$,  which is required just to be  associative and  with isometric  product $\bdot: \cA_{12} \times \cA_{23} \to \cA_{13}$.\par
\medskip
The subsets   of  $\cA = \cA(V, S)$,   defined by
\beq
\begin{split}
&G\= \{x = (x_1, x_2, x_3) + {\bm v} + {\bm s_0} + {\bm s_1}  \in \bR^3 + V + S_0 + S_1\ , \ x_i > 0\}\ ,\\
& G^*\= \{y =  {\bm s_0}^\ast  + {\bm v}^\ast + {\bm s_1}^\ast +  (y_1, y_2, y_3) \in   S'_0 + V' + S_1' + \bR^3 \ , \ y_i > 0\}
\end{split}
\eeq
are closed under the multiplication $\bdot$ and such a product defines  the structure of a simply connected solvable Lie group on each of them.
\begin{definition}
 The group $G$ is called    {\it  Vinberg  triangular group} and $G^*$ is called   {\it dual (triangular) Vinberg  group}.
  \end{definition}
  By the results in  \cite{Vi},  $G$ and $G^*$ are both  simply connected and solvable and
  the map  $\imath: G \to G^*$, $\imath(A) \= (A^*)^{-1}$,  is a Lie group isomorphism.\par
\smallskip
\subsubsection{The  standard matrix representation of a special $T$ algebra}
Let
$$\cA =    (S'_0 +  V' + S_1')+  \bR^3 + (V + S_0 + S_1)$$
be  the  special  T-algebra  $\cA = \cA(V, S)$,  determined by a  Euclidean vector space  $(V, g_V)$ and
a  metric $\bZ_2$-graded $\Cl(V)$-module
$(S = S_0 + S_1, g_S )$.   Each element   $x  = \sum_{i, j = 1}^3 x_{ij}$ of  $\cA =   \sum_{i, j = 1}^3 \cA_{ij}$   is   representable by the  $3 \times 3$ matrix
\beq  \label{V} X(x) = \begin{pmatrix}
x_{11} & x_{12} & x_{13}\\
x_{21} & x_{22} & x_{23}\\
x_{31}& x_{32} & x_{33}
\end{pmatrix} =  \begin{pmatrix}
x_1 &   s_0  & s_1\\
 t_0^* & x_2 &   v\\
t_1^\ast& w^* & x_3
\end{pmatrix}\ .\eeq
 where $x_{i} = x_{ii} \in \cA_{ii} = \bR$   and
\begin{align*}
& x_{12}  = s_0 \in \cA_{12} =S_0\ ,&& x_{13} = s_1 \in \cA_{13} = S_1\ ,&& x_{23} = v \in \cA_{23} =  V\ ,\\
& x_{21} =   t_0^*  \in \cA_{21} = S_0',&& x_{31} = t_1^\ast \in \cA_{31} = S'_1\ ,&& x_{32} = w^*\in \cA_{32} = V^*\ .
\end{align*}
The product $\bdot$ of $\cA$  defines the  (non-associative) product between these matrices,   determined  by the standard matrix multiplication,
$$ (x_{ij}) \bdot (y_{\ell m}) \= 	\bigg( z_{im} 	= \sum_{j = 1}^3 x_{i j} \bdot y_{j m}\bigg)\ .$$
 Since this  matrix product  satisfies
$$X(x) \bdot X(y) = X(x\bdot y)\ ,\qquad \text{for any}\ x, y\in \cA\ ,$$
the linear map
$$\cA \longrightarrow \operatorname{Mat} = \operatorname{Mat}(V,S) \= \{X \= X(x)\ ,\ x \in \cA \}\ ,\quad
 x \longmapsto X(x)\ ,
$$
is a linear representation, called the {\it  (standard)  matrix  representation} of $\cA$. \par
\smallskip
\subsubsection{The space of Hermitian matrices $\ch$ and the  representations of the group  $G$ and $G^*$  in  $\ch$}
\label{dualrepr}
In terms of the    matrix  representation $\operatorname{Mat} =   \operatorname{Mat}(V,S)$,  the  involution  $(\cdot)^\ast$ of the $T$-algebra $\cA$ is given by
\beq X^* = \begin{pmatrix}
x_1 &   s_0  & s_1\\
t_0^\ast  & x_2 &  v \\
t_1^\ast &w^\ast  & x_3
\end{pmatrix}^{\hskip -5pt \mathlarger{\ast}}\=  \begin{pmatrix}
x_1 & t_0  & t_1\\
 s_0^* & x_2 &w \\
s_1^\ast &v^*  & x_3
\end{pmatrix}  \ .\eeq
Using this map, we can express the  scalar product $(\cdot, \cdot)$ of $\cA$ in terms  the matrix representation  by the formula
\beq \label{scalarproduct} ( x, y) = \tr\left( X(x) \bdot Y^*(y)\right)\eeq
The {\it space of Hermitian matrices  in $\operatorname{Mat}$}  is the subspace defined by
\beq \cH \= \{\ X \in \operatorname{Mat}\ :\ X = X^\ast \ \}\ . \eeq
It has  a natural algebra structure determined by   the  Jordan multiplication
\beq \label{defcirc} X \circ Y \= \frac{1}{2} (X Y + Y X)\eeq

\begin{definition} The commutative
algebra  $\cH$
is  called the  {\it Hermitian  Vinberg algebra}  associated with the metric $\Cl(V)$-module $(S, g_S)$.
\end{definition}
\par
In the  matrix representation,  the (upper)  triangular group $G$ and the  dual (lower) triangular group $G^*$
are represented by  the non-degenerate upper  triangular matrices
\beq
\label{matrix:eq}
A =  \begin{pmatrix}
{\bm \r}_1 & {\bm s}_0 & {\bm s}_1\\
0 & {\bm \r}_2   & {\bm v}\\
0 & 0      & {\bm \r}_3
\end{pmatrix}  \ ,\qquad {\bm \r}_i > 0\ ,\  {\bm v} \in V\ , \ {\bm s}_0 \in S_0\ ,\ {\bm s}_1\in S_1\ ,
\eeq
and the lower  triangular matrices
  \beq  \label{matrix:eq-bis}
  B = \begin{pmatrix}
{\bm \t}_1 & 0  & 0 \\
 {\bm t}_0^\ast & {\bm \t}_2   & 0\\
{\bm t}_1^\ast &{\bm w}^\ast & {\bm \t}_3
\end{pmatrix} \ ,\qquad  {\bm \t}_i > 0\ \ , \ {\bm w} \in V\ , \ {\bm t}_0 \in S_0\ ,\ {\bm t}_1\in S_1\ ,
\eeq
respectively. The  isomorphism  $\imath: G \to G^*$, $\imath(A) = A^{-1*}$  is  explicitly given by $$A =  \begin{pmatrix}
{\bm \r}_1 &  {\bm s}_0 & {\bm s}_1\\
0 & {\bm \r}_2   & {\bm v}\\
0 & 0      & {\bm \r}_3
\end{pmatrix} \longrightarrow   A^{-1\ast} = \begin{pmatrix}
\frac{1}{{\bm \r}_1} & 0 & 0   \\
- \frac{{\bm s}_0^\ast}{{\bm \r}_1 {\bm \r}_2}  & \frac{1}{{\bm \r}_2}   & 0 \\
- \frac{  ({\bm \r}_2 {\bm s}_1 - {\bm s}_0\bdot {\bm v})^\ast}{{\bm \rho}_1 {\bm \rho}_2 {\bm \r}_3}  &  - \frac{{\bm v}^*}{{\bm \r}_2 {\bm \r}_3}     & \frac{1}{{\bm \r}_3}
\end{pmatrix} \ .$$
The  Lie algebra $\gg$  (resp.  $\gg^*$)  of the  group  $G$   (resp. $G^*$)  consists   of  all  upper  triangular (resp., lower triangular) matrices in   $\operatorname{Mat} \simeq \cA$.
 The groups of  upper  and lower triangular matrices \eqref{matrix:eq} and \eqref{matrix:eq-bis}  are called {\it standard  realisations} of   $G$ and   $G^*$, respectively.
\par
\smallskip
\subsection{Special Vinberg cones  and their dual cones} \label{specialVinbergcones}
\subsubsection{The   linear actions  of  $G$ and $G^*$   on the space of Hermitian matrices $\cH$}
As we mentioned above,
the Lie algebras
$\gg = Lie(G)$  and $\gg^* = Lie(G^*)$ consist of    arbitrary upper  and lower triangular matrices in $\operatorname{Mat} \simeq \cA$. As it can be directly checked using the axioms of $T$-algebras, the   formula
 \beq  T_B X \= B\bdot X + X\bdot B^*, \qquad  X \in \cH
\eeq
defines  two  linear representations $T: \gg \times \cH  \rightarrow\cH$ and $T: \gg^* \times \cH \rightarrow \cH$ of these Lie algebras on the vector space $\cH$.
Since the Lie groups  $G, G^*$ are solvable and simply connected,  these  Lie algebras representations   integrate to  linear  representations of the groups $G$ and $G^*$
   given by
\beq \label{group-action}
\begin{split}
 \exp (B) (X) &\!\!\= \exp (T_B) X = X + T_BX + \frac12 T_B^2X + \ldots \\
&= X +B\bdot X + X\bdot B^* +\\
& \hskip 0.5 cm +\frac12 \bigg(B\bdot(B\bdot X) + B\bdot (X\bdot B^*) + (B\bdot X)\bdot B^* + (X\bdot B^*)\bdot B^*\bigg) + \dots .
\end{split}
\eeq
Due to the non-associativity of  the algebra $\cA$,  {\it  in general  this  action of the elements $A = \exp (B)$ in $G$ or $G^*$  on the elements $ X \in \cH$ cannot be reduced to  the standard expression
$ A\bdot X\bdot A^\ast$}. \par
\smallskip
\subsubsection{The linear   representation of $G$ on the dual space $\cH' = \Hom(\cH, \bR)$}
Given the vector space of Hermitian matrices   $\cH \subset \operatorname{Mat}$,   we  use  the Euclidean metric  $\langle X,Y\rangle  \= \tr X\bdot Y$ of   $\cH$ to  identify the dual vector space  $\cH' = \Hom(\cH, \bR)$ with $\cH$.  More precisely,
 any $X \in \cH$ is identified with the 1-form  $X^\flat = \langle X, \cdot \rangle$.  \par
 \smallskip The elements $A$ of the group  $G$ act  on  $ \cH'$  by   the dual  transformations
$$A(X^\flat)(Y) \= X^\ast(A^{-1}(Y)) = \tr(X \bdot (A^{-1}(Y))\ , \qquad Y \in \cH\ . $$
We denote by $G'$ the  group  of these dual transformations.  It is the exponential of the dual action of the elements $B$ of the Lie algebra $ \gg = Lie(G)$ given by
\beq  \label{infact} T_B(X^\flat) = (- T_{B^*} (X))^\flat = - (B^*\bdot X + X\bdot B)^\flat\ .\eeq
Indeed,  for any $Y \in \cH$,
\begin{multline} T_B(X^\flat)(Y)  = -  \tr\bigg(X \bdot (B\bdot Y) + X \bdot (Y\bdot B^*)\bigg) \overset{\text{Ax. III \& Ax. IV of  $T$-algebras}}=  \\
= -  \tr((X \bdot B)\bdot Y)) - \tr((B^\ast \bdot X) \bdot Y) = \\
= - \tr(T_{B^*}(X) \bdot Y) = (- T_{B^*}(X))^\flat(Y)\ .
\end{multline}
 Exponentiating  both sides of  \eqref{infact},  it follows that  for any $A  = \exp(B)\in G$,
\beq A(X^\flat) = (\exp(T_B)(X^\flat) = (\exp(T_{-B^*})(X))^\flat = (A^{-1\ast}(X))^\flat\ .\eeq
Hence, under the above  identification  $\cH' \simeq \cH$,  the  group  $G' $ acting on $\cH'$
corresponds to  the group $G^* = \{ \ \imath(A) = A^{-1\ast} \ ,\ A \in G\}$  acting on $\cH$.
\par
\smallskip
\subsubsection{The special  Vinberg cones and their dual and adjoint cones}
By Vinberg's results in \cite{Vi}  the following holds.
\begin{theo}\label{st:thm}
The  orbits $\cV = G(I)$ and $\cV^* = G^*(I)$  of the identity matrix $I \in \cH$  are equal to
\beq
\cV = \{  A \bdot A^*,\, A \in G\}\ ,\qquad
\cV^* =   \{  A^* \bdot A,\, A \in G\}\ .
\eeq
They are both homogeneous convex  cones, on which    the  groups $G$ and, respectively,  $G^*$ act  simply transitively.
\end{theo}
  \begin{definition} The convex cones $\cV = G(I)$   and $\cV^* = G^*(I)$
   are called {\it special Vinberg cone} and {\it its dual cone}, respectively, associated with the  metric $\bZ_2$-graded $\Cl(V)$-module $(S, g_S)$.
\end{definition}
 The dual cone $\cV^*$ has the following important geometric role.  Consider  the
 {\it adjoint  cone}  of the cone $\cV \subset \cH$, that is  the cone in $ \cH' \simeq \cH$ defined by
 $$\cV'=  \{\ X \in \cH\ : \ \tr(X\bdot Y) > 0\ ,\ Y \in \overline{\cV} \setminus \{0\} \ \} .$$
It  can be proved  that
$\cV' = \cV^* = G^*(I)$.
 We finally recall the following
 \begin{definition} \label{self-adjoint} A homogeneous  cone  $\cV \subset \cH$ is called {\it self-adjoint} or {\it symmetric} if there exists a vector space isomorphism $L: \cH \to \cH$ such that $L(\cV)  =  \cV^* $.
 \end{definition}
 \par
 \smallskip
\subsection{Group coordinates, adapted orthogonal coordinates and De Wit and Van Proeyen coordinates}
 Since both cones  $\cV = G(I)  $ and  $\cV^* =  G^*(I) $ are    in natural bijection with  $G$, we may consider  the  diffeomorphisms
$\xi_G: \cV \to G$ and $\xi^*_G: \cV^* \to G$  given  by
$$X = A \bdot A^*  \overset{\xi_G} \longmapsto A\ ,\qquad Y = A^{-1*} \bdot A^{-1}  \overset{\xi^*_G} \longmapsto A\ .$$
We call them  the {\it group coordinates of $\cV$ and $\cV^*$}.
 \par
The   relations between the    group coordinates and the corresponding elements in  $\cV$ and $\cV^*$ are as follows.
If $X \in \cV$  corresponds to the element  $  A =  \bigg(\smallmatrix
{\bm \r}_1 & {\bf s}_0 & {\bm s}_1\\
0 & {\bm \r}_2   &{\bm v}\\
0 & 0      & {\bm \r}_3
\endsmallmatrix \bigg) \in G$,  than the entries of
$X =     \bigg(\smallmatrix
x_1 &s_0   & s_1\\
 s_0^\flat & x_2 &  v\\
s_1^\flat &v^\flat & x_3
\endsmallmatrix\bigg) =  A \bdot A^*$
are
 \begin{align} 
\nonumber &x_1 = {\bm \r}_1^2 + |{\bm s_0}|^2  + |{\bm s}_1|^2\ ,&&
x_2 = {\bm \r}_2^2+|{\bm v}|^2 \ , &&
x_3 =   {\bm \r}_3^2\ , \\
\label{Hgg2} & {\bm s}_0 + {\bm s}_1\bdot {\bm v}^\flat\ ,&&
v =   {\bm \r}_3 {\bm v}\ ,&&s_1 ={\bm \r}_3 {\bm s}_1\ .
\end{align}
Similarly, if  $X\in \cV^*$ corresponds to $\xi_G^*(X)  =  A =  \bigg(\smallmatrix
{\bm \r}_1 & {\bm s}_0& {\bm s}_1\\
0 & {\bm \r}_2   &{\bm v}\\
0 & 0      & {\bm \r}_3
\endsmallmatrix \bigg)$,   the entries of $X = A^{-1} {}^* \bdot A^{-1} $ are
\begin{align} 
\nonumber &x_1 =  \frac{1}{{\bm \r}_1^2}\ ,\hskip 1 cm
x_2 =  \frac{({\bm \r}_1{\bm \r}_3)^2+ {\bm \r}_3^2| {\bm s}_0|^2}{({\bm \r}_1 {\bm \r}_2 {\bm \r}_3)^2} \ ,  \\
\nonumber &\hskip 5 cm x_3 =  \frac{ |{\bm \r}_2 {\bm s}_1 -{\bm s}_0\bdot {\bm v}|^2 + {\bm \r}_1^2 |{\bm v}|^2 + {\bm \r}^2_1  {\bm \r}^2_2}{({\bm \r}_1 {\bm \r}_2 {\bm \r}_3)^2} \ ,  \\
\label{Hgg2bis} &s_0 =  -\frac{{\bm \r}_2 {\bm \r}^2_3{\bm s}_0 }{({\bm \r}_1{\bm \r}_2 {\bm \r}_3)^2} \ ,\hskip 1 cm
v =   \frac{ {\bm \r}_3   {\bm s}_0^\flat \bdot ({\bm \r}_2 {\bm s}_1 - {\bm s}_0 \bdot {\bm v}) + {\bm \r}_1^2 {\bm \r}_3 {\bm v}}{({\bm \r}_1 {\bm \r}_2 {\bm \r}_3)^2} \ ,
\\
\nonumber &\hskip 8 cm s_1 =  \frac{- {\bm \r}_2 {\bm \r}_3 ({\bm \r}_2 {\bm s}_1 -  {\bm s}_0 \bdot {\bm v})}{({\bm \r}_1 {\bm \r}_2 {\bm \r}_3)^2}\ .
\end{align}
The  inverses to the  \eqref{Hgg2}
are given by (see also \cite{Vi, AC})
\begin{align} \label{Vinverses}
\nonumber &{\bm \r}_3^2  =  x_3\ ,   \\
\nonumber &  {\bm \r}_2^2  = x_2-  \frac{1}{{\bm \r}_3^2}|  v|^2 =   \frac{x_3 x_2-  |   v|^2 }{x_3}\ , \\
\nonumber & {\bm \r}_1^2 = \frac{   x_1x_2x_3  -  x_1|  v|^2 - x_2 |s_1|^2 -  x_3  |  s_0 |^2   + 2 (s_0,  s_1\bdot v^\flat)}{ ({\bm \r}_2 {\bm \r}_3)^2} =
\\\nonumber & \hskip 1 cm   =   \frac{ x_1x_2x_3  -  x_1  | v|^2 - x_2   |s_1|^2 -  x_3  |  s_0 |^2   + 2\langle \mu(v, s_0), s_1\rangle}{ x_3 x_2-  | v|^2} \ ,
 \\
  &  {\bm s}_0 = \frac{1}{{\bm \r}_2}s_0   - \frac{1}{ {\bm \r}_2   {\bm \r}_3^2}   s_1\bdot v^\flat \  ,\qquad {\bm v} = \frac{1}{ {\bm \r}_3} v\ ,\qquad   {\bm s}_1 = \frac{1}{ {\bm \r}_3} s_1\ ,
\end{align}
while the    inverses   to the  \eqref{Hgg2bis}  are  given  by
\begin{align} \label{Vinverses*}
\nonumber & \frac{1}{{\bm \r}_1 ^2} =  x_1\ ,\\
 \nonumber & \frac{1}{ {\bm \r}_2^2}     = x_2-   {\bm \r}_1^2| s_0|^2 = \frac{x_1 x_2 -      | s_0|^2}{x_1} \ ,  \\
\nonumber &  \frac{1} {{\bm \r}_3^2} =   ({\bm \r}_1 {\bm \r}_2)^2 \bigg( x_1x_2x_3  -  x_1|  v |^2 - x_2 |s_1|^2 -  x_3  | s_0 |^2   + 2  \langle s_0^\flat \bdot s_1 ,  v\rangle + \gd \bigg)  =
\\
\nonumber & \hskip 1 cm   = \frac{  x_1 x_2  x_3   -    x_1 |    v|^2  - x_2|  s_1|^2  -     x_3 |s_0|^2 + 2      \langle \mu(v, s_0), s_1\rangle +  \gd}{  x_1 x_2 -      | s_0|^2}\ ,\\
\nonumber & \hskip 4 cm \text{where}\ \gd \= \frac{ |s_0|^2 |s_1|^2- \langle s_1\bdot s_0^\flat, s_1 \bdot s_0^\flat\rangle }{x_1}
 \\
\nonumber  &  {\bm s_0}  = - {\bm \r}^2_1 {\bm \r}_2 s_0 \ ,\hskip 1 cm   {\bm v} =   -   {\bm \r}^2_1 {\bm \r}_2^2 {\bm \r}_3  s_0^\flat  \bdot   s_1  +   {\bm \r}_2^2 {\bm \r}_3 v
 \ ,
\\
 &\hskip 5 cm  {\bm \r}_2 {\bm s}_1 = {\bm s}_0 \bdot {\bm v}    -  \frac{({\bm \r}_1 {\bm \r}_2 {\bm \r}_3)^2}{{\bm \r}_2 {\bm \r}_3} s_1  \ .
\end{align}
Note that \eqref{Vinverses}  imply that  $\cV$ (resp.  \eqref{Vinverses*} imply that  $\cV^*$)  coincides with   the  convex set  characterised by the following   three
inequalities
\begin{align}
\nonumber & x_3 > 0\ ,\qquad  x_3 x_2-  |v|^2 > 0\ ,\\
 \label{positivity}
 &   x_1x_2x_3  -  x_1  |  v|^2 - x_2   |s_1|^2 -  x_3  | s_0 |^2   + 2\langle \mu(v, s_0), s_1\rangle > 0\ ,\\
\nonumber  \big(\text{resp.} \ \  & x_1 > 0\ , \qquad  x_1 x_2 -      | s_0|^2> 0\\
&  x_1 x_2  x_3   -    x_1 |   v|^2  - x_2|  s_1|^2  -     x_3 |s_0|^2 + 2      \langle \mu(v, s_0), s_1\rangle + \gd > 0\big)\ ,
\end{align}
which    generalise   Sylvester's criterion for  positive definiteness.
\par
\medskip
We now observe that     $\cH$   is   isomorphic (as a vector space) to  the vector space
$W = \bR^3 + V + S_0 + S_1 = \cA_{11} + \cA_{22} + \cA_{33} +V + S_0 + S_1$.
Therefore each  basis $\cB$ for $W$    naturally determines an associated system of   coordinates for $\cH \simeq W$, which we call {\it standard}.
If  $\cB$ has the form
\beq \cB  = ({\bm 1}_1, {\bm 1}_2, {\bm 1}_3, e_j,   f_{0|\a}, f_{1|\b}, f^*_{0|\a}, f^*_{1|\b} ) \ ,\label{adapted}\eeq
 with ${\bm 1}_i$ standard basis of $\cA_{ii} = \bR$ and $(e_j)$,  $(f_{0|\a}, f_{1|\b})$ orthonormal bases of $(V, g_V)$ and $(S = S_0 + S_1, g_S)$, respectively, the corresponding    coordinates on $\cH$
\beq  \label{118}
X =  \begin{pmatrix}
y^1 {\bm 1}_1 = x_1& s^{\a}_0 f_{0|\a} & s^\b_1 f_{1|\b}\\
  (s^{\a}_0 f_{0|\a})^* & y^2 {\bm 1}_2 = x_2& v^i e_i \\
(s^{ \b}_1 f_{1|\b})^*&  (v^i e_i)^* & y^3 {\bm 1}_3 = x_3
\end{pmatrix} \overset{\cB}\longrightarrow \begin{pmatrix} y {=} (y^i) \\ v {=} (v^j) \\ s_0 {=} (s^{\a}_0)\\ s_1 {=} (s^{\b}_1)\end{pmatrix}\eeq
are called  {\it   adapted orthogonal  coordinates} on $\cH$.
A  basis $\cB'$ for the space $W' = \Hom(\cW, \bR)$, which is dual  to a basis $\cB$ for $W$,  determines coordinates for $\cH' \simeq W'$, which we call {\it dual coordinates} associated with $\cB$.\par
\smallskip
For  a given system of adapted orthogonal   coordinates
$$\bigg(y^1, y^2, y^3,  v = (v^j), s_0 =(s_0^{\a}), s_1 = (s_1^{\b})\bigg)\ ,$$
 the     {\it  associated De\ Wit-Van\ Proeyen   coordinates} are the   coordinates $w = (w^I)$   determined by the linear transformation rules
 \beq \label{transf}
\begin{split}
& w^1 = y^1 \ ,\quad w^2 =   \frac{y^3 + y^2}{2}  \ ,\quad w^3 =  \frac{- y^3 + y^2}{2} \ ,\  w^{\wt \mu} =  v^{\wt \mu - 3}\ ,\\
& w^i =  s_0^{i - (3 + n) }\ ,\ \   w^{i'} = s_1^{i' -(3 + n + \frac{n_S}{2})}\ ,
\end{split}
\eeq
with $4 \leq \wt \mu \leq n + 3$, $4 + n \leq i \leq  3  + n +\frac{ n_S}{2}$ and $4 + n + \frac{ n_S}{2}  \leq i' \leq 3  + n + n_S$,
where  $n \=  \dim V$ and $n_S =\ \dim S {=} \dim S_0 + \dim S_1 {=} 2 \dim S_0$.
The corresponding {\it dual} transformation rules
\beq \label{dual-transf}
\begin{split}
&  w_1 =  y_1\ ,\ \   w_2 =y_3 + y_2\ ,\ \   w_3 = - y_3 + y_2 \ , \ \  w_{\wt \mu} = v_{\wt \mu - 3} ,\\
& w_i = s_{0|i - (3 + n) }\ ,\ \  w_{i'} = s_{1|i' -(3 + n + \frac{n_S}{2})}\ ,
\end{split}
\eeq
define the   {\it  De\ Wit-Van\ Proeyen  dual  coordinates}.
\par
\smallskip
\subsection{The invariant  and dual invariant  cubic  polynomials}
The Lie group $G$ of upper triangular matrices  is the direct product $G = (\bR_+ I) \times G_0$ of the {\it dilatation subgroup}
$\bR_+ I $
and the {\it unimodular subgroup}
$$G_0 \= \left\{\ A = \begin{pmatrix}
{\bm \r}_1 & {\bm s}_0 & {\bm s}_1\\
0 & {\bm \r}_2   & {\bm v}\\
0 & 0      & {\bm \r}_3
\end{pmatrix} \ ,\   {\bm \r}_1{\bm \r}_2{\bm \r}_3  = 1\ \right\} \ .$$
Similarly  $G^*$ is the direct product $G^* = (\bR_+ I) \times G^{\ast}_0$ of the {\it dilatation subgroup}
$\bR_+ I $
and the {\it unimodular subgroup}  $G^{\ast}_0 \subset G^* = \imath(G)$
$$G^{\ast}_0 \= \left\{\ B = \begin{pmatrix}
{\bm \t}_1 & 0  & 0 \\
 {\bm t}_0^\flat & {\bm \t}_2   & 0\\
{\bm t}_1^\flat  &  {\bm u}_0^\flat  & {\bm \t}_3
\end{pmatrix} \ ,\  {\bm \t}_1 {\bm \t_2} {\bm \t_3} = 1\ \  \right\} = \imath(G_0)  \ . $$
\par
\medskip
\begin{definition} \label{invariant-cubic-polynomial}  A non-zero rational function    $p: \cH \to \bR$ is called   {\it invariant cubic} (resp. {\it dual invariant cubic rational function}) if  it is a polynomial (resp. a homogeneous rational map) and satisfies
\beq\label{inv-1} p\big(t A(X)\big) = t^3 p(X)\qquad \text{with} \ t \in \bR_+\ ,\ A \in G_0\ \  (\text{resp.} \ A \in G^*_0\ )\ .\eeq
\end{definition}

\begin{theo} \label{lemma17} Up to a scaling factor, there exist a unique invariant  cubic polynomial   $d: \cH \to \bR$ and a unique dual invariant cubic   rational function
$d^*: \cH \to \bR$.  Assuming the normalising condition $d(I) = d^*(I) = 1$, they are  given by
\beq \label{cubic-true} 
\begin{split}
&  d(X)  =   x_1x_2x_3  -  x_1  |   v|^2 - x_2   |s_1|^2 -  x_3  |s_0 |^2   + 2\langle \mu(v, s_0), s_1\rangle \\
 &d^*(X) =  d(X) + \gd(X)  \ ,\qquad \text{where} \ \gd(X) \= \frac{ |s_1|^2 |s_0|^2- \langle s_1\bdot s_0^\flat, s_1\bdot s_0^\flat \rangle}{x_1}\ \ .
 \end{split}
 \eeq
\end{theo}
\begin{pf}   The invariant  cubic polynomials    and the dual invariant cubic rational functions are  unique up to a scaling because the subgroups $G_0$ and $G_0^*$ have codimension one orbits.
We now prove that, if $d(I) = d^*(I) = 1$, then the formulas for  $d(X)$ and $d^*(X)$ are given by   \eqref{cubic-true}. By real analyticity, it suffices to prove this for  $X  = A \bdot A^* \in G(I)$ and  $X^* = A^* \bdot A \in G^*(I)$
   for some $ A =  \bigg(\smallmatrix
{\bm \r}_1 & {\bm s}_0 & {\bm s}_1\\
0 & {\bm \r}_2   & {\bm v}\\
0 & 0      & {\bm \r}_3
\endsmallmatrix \bigg)$. \par
The squared determinant map on $G$ and the inverse of the squared determinant map on $G^*$, i.e. the functions
$$(\det A)^2 \= ({\bm \r}^1 {\bm \r}^2  {\bm \r}^3)^2 = (a_{11}a_{22} a_{33})^2, \  \frac{1}{(\det A)^2} \= \frac{1}{({\bm \r}^1 {\bm \r}^2  {\bm \r}^3)^2} = \frac{1}{(a_{11}a_{22} a_{33})^2},$$
are $G_0$-invariant and $G^*_0$-invariant, respectively, and they are both equal to $1$ on $A = I$.
They  are therefore  the unique  $G_0$- and $G_0^*$-invariant functions  $d: \cV \to \bR$ and  $d^*:\cV^* \to \bR$ satisfying the normalising condition.  The inverse maps  \eqref{Vinverses} and \eqref{Vinverses*}  show that the expressions of these two functions in terms of the entries of the matrix $X$ are   given by  \eqref{cubic-true}. 
\end{pf}
In   adapted orthogonal coordinates
 the invariant   rational functions \eqref{cubic-true}  are  
%
%
\begin{align} 
\nonumber & d(y, v, s_0, s_1)  = y^1 y^2 y^3  -  y^1  |   v|^2 - y^2   |s_1|^2 -  y^3  |  s_0 |^2   + 2  \g_{i \a \b} v^i s_0^\a s^\b_1  \\
\nonumber & d^*(y, v, s_0, s_1) = \\
\nonumber & \hskip 0.5 cm = y^1y^2 y^3 - y^1 |v |^2-  y^2 |s_1|^2- y^3|s_0 |^2 + 2  \g_{i \a \b} v^i s_0^\a s^\b_1  + \gd(y_1, s_0, s_1) = \\
&  \label{cubic} \hskip 0.5 cm  = d(y, v, s_0, s_1)   + \gd(y_1, s_0, s_1)\ \\
\nonumber & \hskip 1 cm  \text{where } \qquad \gd(y_1, s_0, s_1) \= \frac{ |s_0| ^2 |s_1|^2 - \sum_{i = 1}^n\left(\g_{i \a \b} s_0^\a s^\b_1\right)^2}{y_1}
\end{align}
and where  $\g_{i \a \b}$ is the   $\G$-matrix representing   the Clifford multiplication $\mu(e_i, \cdot)$.
Let us now denote by $\widecheck \g_3 = (\widecheck \g_{3{\bf i}{\bf j}})$ and $\widecheck \g_{\wt \mu} = (\widecheck \g_{\wt \mu {\bf i}{\bf j}})$ the square matrices,  acting on the  $n_S$-dimensional  space  $S = S_0 + S_1$ (whose elements have components  denoted by   $(s^{\bf i}) = (s_0^\a, s_1^{\b})$ for short) given by
$${\widecheck \g}_{3}= \left(\begin{array}{cc} I_{\frac{n_S}{2}} & 0 \\ 0 & - I_{\frac{n_S}{2}}   \end{array}  \right) \ , \qquad {\widecheck \g_{\wt \mu}} \= \left(\begin{array}{cc} 0  &\g_{\wt \mu - 3}\\ - \g_{\wt \mu-3 } & 0 \end{array} \right)\ .$$
By  the transformation rules \eqref{transf},  the expression for $d $  in  de Wit - Van Proeyen coordinates $ w = (w^I) = (w^1, w^2, w^3, w^{\wt \mu}, w^{\bf i})$  becomes
\begin{multline} \label{dWvP}
 d(w^I) = \\=   w^1 (w^2)^2  - w^1 (w^3)^2 - w^1 \d_{\wt \mu \wt \nu} w^{\wt \mu} w^{\wt \nu} -  w^2  \d_{i j}  w^{\bf i} w^{\bf j}  +  {\widecheck \g}_{3 {\bf i}{\bf j}} w^3 w^{\bf i} w^{\bf j} +  {\widecheck \g}_{\wt \mu {\bf i} {\bf j}} w^{\wt \mu} w^{\bf i} w^{\bf j} = \\
=   w^1 (w^2)^2  - w^1 \d_{\mu \nu} w^\mu w^\nu -  w^2  \d_{{\bf i} {\bf j}}  w^{\bf i} w^{\bf j}  +  {\widecheck \g}_{ \mu {\bf i} {\bf j}} w^{ \mu} w^{\bf i} w^{\bf j}  \ .
 \end{multline}
 where   $\mu$ runs   between $3$ and $3 + n$ and ${\bf i}, {\bf j}$ run between $4 + n$ and $3 + n + n_S$ (\footnote{Mind the differences in  indices conventions: on one hand in
 adapted orthogonal coordinates we denote the vector (resp. spinor) indices  by Latin letters as $i, j$ (resp. Greek letters like $\a$, $\b$); on the other hand,  following a traditional choice of physics literature, in
 the de Wit - van Proeyen coordinates  the vector (resp. spinor)  indices are denoted by  Greek letters of the sequence  $\mu$, $\nu$, etc. (resp. Latin  letters   ${\bf i}$, ${\bf j}$ and so on).}).  Note that
    \begin{itemize}[leftmargin = 15pt]
\item[--] the $\widecheck \g_\mu$   are the Dirac matrices of    the representation of the Clifford algebra $\cC \ell(\wt V, g_{\wt V})$, $\wt V \= \bR \oplus V$, $g_{\wt V} \=   dt^2 \oplus  g_V$, on the spinor space $S = S_0 + S_1$;
   \item[--] up to the  factor $3$,   \eqref{dWvP}  is  the formula found  by  De Wit and Van Proeyen
in their   classification of   invariant cubic polynomials  (\cite[Formula (4.2)]{WP1}).
\end{itemize}
This motivates the name of the coordinates $(w^I)$,
one of  the most  frequently used set of coordinates in the physics literature on invariant cubic polynomials.
\par
\medskip
A non-zero rational function     $p': \cH' \to \bR$    is called {\it  invariant  dual cubic rational function}  if
  $p'\big(t A' (X)\big) = t^3 p'(X)$ for any $t A \in G = (\bR_+ I) \times G_0$.  \par
  \smallskip
The   isomorphism $(\cdot)^\flat: \cH \to \cH'$ maps  each  invariant cubic  rational function  $d^* = d + \gd$  on  $\mathcal{H}$  onto a uniquely associated invariant dual cubic  rational function  $d' $ $+ \gd'$ on $ \mathcal{H}’$.
Thus,  Theorem \ref{lemma17}
   implies that {\it  up to  a scaling, there is a unique invariant dual cubic rational function  $d'   + \gd': \cH' \to \bR$, which is determined as follows. Given an adapted system of coordinates $x = (x^a)$ on $\bR^n = \cH$,  if  we denote by $g= (g_{o ab})$  the  components  of  the Euclidean scalar product $\langle \cdot, \cdot \rangle$ in the coordinate basis,
 by  $d_{abc}$  the (symmetric in all indices) coefficients  of the   cubic polynomial $ d(x) = d_{abc} x^a x^b x^c$ and  by  $g^{-1}_o \= (g_{o}^{ab})$  the inverse of $g_o$, then   $d'(x)+ \gd'(x) = g_o^{a a'} g_o^{b b'} g_o^{c c'} d_{a' b' c'} x_a x_b x_c$  $ +  \gd(g_o^{ab} x_b)$. }\par
 This means that,  if the coordinates are  orthogonal (thus with $g_o = (g_{o ab}) = (\d_{ab})$) and the components of  $d$ in these coordinates are denoted by $d_{abc}$, then  the components  $d^{abc}$ of   the dual polynomial $d'$ in the associated
  dual   adapted orthogonal   coordinates are  obtained by simply raising the  indices with the matrix $\d^{ab}$.  In particular, if the coordinates are not just orthogonal, but also {\it adapted}  (i.e. as in \eqref{118}), then the expression for $d'$ $+\gd'$  is 
 \begin{multline}\label{thed'}
(d' + \gd')(y, v^\flat, s^\flat_0, s^\flat_1) = y_1 y_2 y_3 - y_1 | v|^2 -  y_2 |s^\flat_1|^2 - y_3|s_0^\flat|^2 + 2 \g^{i \a \b} v_i s_{0\a} s_{1\b} +
\\
  + \frac{|s_0| ^2 |s_1|^2 - \sum_{i = 1}^n \left(\g^{i \a \b} s_{0\a} s_{1\b}\right)^2 }{y_1} \ .
\end{multline}
 On the other hand,  since the De Wit - Van Proeyen    coordinates $(w^I)$ are obtained from the adapted orthogonal coordinates  $(y^a, v^j, s_0^\a, s_1^\b)$  by means of the  {\it non-orthogonal} transformation \eqref{transf},  the entries of the  matrix  $(g_{o I J})$   and  of its inverse in such new coordinates  are
\beq \label{237} \left( g_{o I J}\right) = \left(\begin{array}{ccccc} 1 & 0 & 0 & 0\\ 0 & 2  & 0 & 0\\ 0 & 0 & 2 & 0\\ 0 & 0 & 0 & I_{n + n_S}\end{array} \right)\ ,\qquad  \left( g_o^{I J}\right) = \left(\begin{array}{ccccc} 1  & 0 & 0 & 0\\ 0 & \frac{1}{2} & 0 & 0\\ 0 & 0 & \frac{1}{2} & 0\\ 0 & 0 & 0 & I_{n + n_S}\end{array} \right)\ .\eeq
Hence  the expression  for $d'   + \gd'$ in the De Wit - Van Proeyen coordinates is
\begin{multline}  \label{dWvP-prime} (d'+  \gd')(w_I) =   \frac{1}{4}  w_1 w_2^2  - \frac{1}{4}  w_1 w_3^2 - w_1 \d^{\wt \mu \wt \nu} w_{\wt \mu} w_{\wt \nu} - \frac{1}{2} w_2  \d^{{\bf i}{\bf j}}  w_{\bf i} w_{\bf j}  +\\
+  \frac{1}{2}  {\widecheck \g}_{3}^ {\phantom{3}{\bf i}{\bf j}} w_3 w_{\bf i} w_{\bf j} +  {\widecheck \g}^{\wt \mu {\bf i} {\bf j}} w_{\wt \mu} w_{\bf i} w_{\bf j} + \\
  + \frac{ \left(\sum_{{\bf i} = 4 +n}^{3 +n + \frac{n_S}{2}}  w_{\bf i}^2\right) \left(\sum_{{\bf j} = 4 +n + \frac{n_S}{2}}^{3 +n + n_S}  w^2_{\bf j}\right)
- \sum_{\wt \mu = 1}^n\left(\g^{\wt \mu {\bf i}{\bf j}} w_{{\bf i}} w_{{\bf j}}\right)^2}{y_1}
\  ,
\end{multline}
where $\wt \mu$ runs between $4$ and $3 + n$ and ${\bf i}, {\bf j}$ run between $4 + n$ and $3 + n + n_S$.
\par
\smallskip
\subsection{Quadratic maps associated with  invariant  cubic polynomials  and cubic rational functions, and their  inverses}
In what follows, we identify any cubic polynomial  $d: \cH \to \bR$  and any dual   cubic polynomial  $d': \cH' \to \bR$
with the associated  symmetric tenors
$d\in S^3  \cH'$ and $d' \in S^3 ( \cH')' = S^3 \cH$ determined  by polarisation. \par
 Given a special Vinberg cone $\cV \subset \cH$,  we denote by $d_\cV$ and $d'_\cV + \gd_\cV'$  the uniquely associated invariant  cubic and  invariant dual   cubic  rational function, defined by \eqref{cubic-true} and \eqref{thed'}, respectively,  satisfying the normalising conditions $d_\cV(I) = 1$ and $(d'_\cV+ \gd'_\cV)(I^\flat) = 1$.  
  By Theorem \ref{lemma17}
all other invariant cubic   polynomials and invariant  dual  cubic rational functions associated with $\cV$ have the forms
 \beq d = k d_\cV \quad \text{with}\ k \= d(I)\ ,\qquad d'   + \gd'= k' d'_\cV   + k' \gd'_{\cV}\quad \text{with}\ k' \=  (d' + \gd')(I^\flat)\ .\eeq
 With no loss of generality from now on we assume   $k  >0$ and $k' > 0$.  Actually, given  $d = k d_\cV$, we constantly denote by  $d'+ \gd'$ the  {\it canonically associated    dual   rational function}, which is  defined by
 $$ d' + \gd'\= \frac{1}{k} d'_{\cV}  + \frac{1}{k} \gd'_{\cV}\ .$$
 \par
As we announced in the introduction,
the  problem of determining the entropy of  extremal BPS black holes in the supergravity theories  of this paper
can be  reduced to the  mathematical question  of finding    the inverses of   certain  quadratic maps  determined by   cubic polynomials (see \cite{GMR}).
In the  cases in which the   scalar manifold   is a homogeneous space determined by an irreducible cubic polynomial, such a   cubic polynomial  is the  invariant  polynomial $d$ associated with an appropriate special Vinberg cone. In these cases, the    quadratic map  that needs  to be inverted  is defined as follows.
\begin{definition} Given a   invariant cubic polynomial  $d = k d_\cV$  and its canonically associated  dual  cubic rational function $d'  + \gd'= \frac{1}{k} d'_\cV + \frac{1}{k} \gd'_{\cV}$,
 the  {\it associated quadratic maps}  are  the maps
$h: \cH \to \cH' = \Hom(\cH, \bR) $  and $h' + \gh': \cH'  \to  \cH = \Hom(\cH', \bR)$ defined by
\beq\begin{split}
& h(X)(Y)  \= \frac{1}{3} \frac{d}{ds} \left(d(X + s Y)\right)\big|_{s = 0} = d(X, X, Y) \ ,\\
&
(h' + \gh')(X^\flat)(Y^\flat)  \= \frac{1}{3}\frac{d}{ds} \left((d'+ \gd')(X^\flat + s Y^\flat)\right)\big|_{s = 0}  =\\
& \hskip 2 cm  =   d'(X^\flat, X^\flat, Y^\flat) +  \gh'(X^\flat)(Y^\flat)\ ,\\
& \hskip 3 cm \text{with}\ \  \gh'(X^\flat)(Y^\flat) \= \frac{1}{3}\frac{d}{ds} \left(\gd'(X^\flat + s Y^\flat)\right)\big|_{s = 0}  \ .
\end{split}
\eeq
\end{definition}
The next lemma and theorem  solve the mentioned  inversion problem and  leads  to the solution to
 our BPS black hole entropy   problem under the assumption that the scalar manifold is homogeneous, but not necessarily symmetric (see \S\  \ref{conclusion}).   \par
\medskip
\begin{lem} \label{lemma29}  Let $\cV  \subset \cH$ be a special Vinberg cone with associated  $d = k d_\cV$,  $d'  + \gd' = \frac{1}{k} (d'_\cV + \gd'_\cV)$ and with  $h$, $h'  + \gh'$   corresponding quadratic maps, as defined  above.   Let also denote 
\beq \nonumber \begin{split}
&\cH_+ \= \{d > 0\ ,\ y_1 \circ h \neq 0\} , \  \cH_- \= \{d < 0 \ , \ y_1 \circ h\neq 0\}\ \ \subset \cH\ ,\\
&  \cH'_+ \= \{d'  + \gd' > 0\ ,\ y_1 \neq 0\} , \ \ \ \ \ \ \ 
 \cH'_- \= \{d' + \gd' < 0\ ,\  y_1 \neq 0\} \ \subset \cH'\ .
\end{split}
\eeq
\begin{itemize}[leftmargin = 20pt]
\item[(i)] The maps $h$ and $h'  + \gh'$ are $G_0$-equivariant with respect to the  natural $G_0$-actions on $\cH$ and $\cH' = \Hom(\cH, \bR)$, i.e.
\beq
h\big(A(X)\big) =  A'\big(h(X))\qquad\text{and}\qquad   (h' + \gh' )\big(A'(X^\flat)\big) = A\big((h' + \gh')(X^\flat)\big)\ .
\eeq
Hence $h$ and $h'$  map any $G$-orbit in $\cH$ into a $G$-orbit in $\cH'$.
\item[(ii)]   $h(I) = k I^\flat$ and $(h' + \gh')(I^\flat) = \frac{1}{k} I$.
\item[(iii)] for any $X \in \cH$  with $x_1(X) \neq 0$
\beq \label{formula}  ((h'   + \gh')\circ h)(X) = d(X) X\  \qquad  (h\circ (h'+  \gh'))(X^\flat) =  \big(d'(X^\flat)+ \gd'(X^\flat)\big) X^\flat\ .\eeq
\item[(iv)]  The map   $h$ (resp. $h' + \gh'$)  determines
 a $G_0$-equivariant diffeomorphism from each $G$-orbit $\cV = G(I)$ in $\cH_+ \cup \cH_-$ (resp. $\cH'_+ \cup \cH'_-$) into  a  $G$-orbit in $\cH_+'$
 (resp. in $\cH_+$) and it maps
     $\cH_0 \= \{ d = 0\ , x_1 \circ h\neq 0\}$ into $ \cH'_0 \= \{d' = 0 \ ,\ x_1  \neq 0\}$  (resp. $\cH'_0$ into $\cH_0$); in particular  the maps
$h: \cV \to \cV'$ and $h': \cV' \to \cV$ are  diffeomorphisms.
\item[(v)]  Each map $h: \cH_\pm \to  \cH'_+ $ (resp. $h'  + \gh': \cH'_\pm \to  \cH_+$)   is a  diffeomorphism  onto $\cH'_+   \subset \cH'$
(resp.  onto $\cH_+   \subset \cH$).
\end{itemize}
\end{lem}\begin{pf}   (i) For the first identity  it suffices to observe that
\begin{multline*} h\big(A(X)\big)(Y) =  \frac{1}{3} \frac{d}{ds} \left(d(A(X + s A^{-1}Y))\right)\big|_{s = 0} =  \\
=  \frac{1}{3} \frac{d}{ds} \left(d(X + s A^{-1}Y)\right)\big|_{s = 0} 
= A'\big(h(X)\big)(Y)\ .
\end{multline*}
 The proof of the second identity is similar.\\
  (ii)  It is a direct consequence of
\begin{align*} 
& h(I)(Y)  =  \frac{k }{3  } \frac{d}{ds } \big(d_\cV (I + s Y)\big)\big|_{s = 0}  = k\Tr(Y) = k  I^\flat(Y)\ ,\\
& (h' + \gh')(I^\flat)(Y^\flat)  =  \frac{1}{3k } \frac{d}{ds } \big( (d'_\cV + \gd_\cV')(I^\flat + s Y^\flat)\big) \big|_{s = 0}  =  \frac{1}{k} \Tr(Y^\flat) =  \frac{1}{k} I(Y^\flat)
.\end{align*}
  (iii) We first prove the first identity of  \eqref{formula} in case $X \in \cV$, i.e. under the assumption that  $X = t A(I)$
  for some   $t \in \bR_+$ and $A \in G_0$. Then
\begin{multline*} \big((h'  + \gh') \circ h\big)(X) = (h' + \gh') \big(h\big(t A (I)\big)\big) {=} \\
=  t^4 (h'+  \gh')\big(  k A'(I^\flat)\big) {=} k^2 t^3 (t A (h'+  \gh')(I^\flat)) {=} t^3 k (t A(I) )  {=} d( X) X. \end{multline*}
A similar argument proves the second identity in case  $X^\flat \in \cV'$. Since both sides of  the two identities  of \eqref{formula}  are rational functions,  it follows that they  hold  for any $X \in \cH$ for which both sides are well defined.  \\
(iv)  Let  $d(X) \neq 0$ and  $X' \= h(X)$. By \eqref{formula} we have   $(h'  + \gh')(X') = d(X) X$ and therefore
\beq \label{formula-bis}
\begin{split}&   (d'  + \gd')(X')  =   \frac{1}{7} \left(   (d'  + \gd')(2 X')  -  (d' + \gd')(X') \right) =\\
&  \hskip 0.2 cm  = \frac{1}{7} \int_0^1 \frac{ d \big((d' + \gd')(X' + t X') 
)}{ d t} \bigg|_{t = \s} d \s =\\ 
& \hskip 0.2 cm = \frac{3}{7} \int_0^1\frac{1}{3} \frac{ d \big((d' + \gd')(X' + \s X' + s X') 
)}{ d s} \bigg|_{s = 0} d \s = \\
&\hskip 0.2cm = \frac{3}{7} \int_0^1  (h' + \gh')((1+ \s)X')(X') d \s  = (h' + \gh')( X')(X') \left[ \frac{3}{7} \int_0^1 (1 +\s)^2 d \s \right]=     \\
& \hskip 0.2cm  =  (h' + \gh')(X')(X')= d(X) X'(X) = d(X) h(X)(X) = d(X)^2 > 0\ .\\
\end{split}
\eeq
This shows that  $h( \cH_\pm )  \subset \cH'_+$ and, by a similar argument, that   $(h' + \gh')(\cH'_\pm)  \subset  \cH_+ $. The claim then follows from  (i) and (iii).\par
(v) For the maps $h: \cH_\pm \to \cH'_+$,  it suffices to observe that for any $h(X) \in \cH'_+$ with $X \in \cH_\pm$
\beq \label{inv}   \frac{1}{\sqrt{(d'  + \gd')(h(X))}} (h' + \gh')(h(X)) \overset{\eqref{formula} \& \eqref{formula-bis}  } = \frac{1}{\sqrt{(d(X))^2}}  d(X) X =\pm X \ . \eeq
 A similar argument shows the existence of the inverse for $h' + \gh': \cH'_{\pm} \to \cH_+$.
\end{pf}
The previous lemma   and  \eqref{inv}   imply the following
\begin{theo} \label{thetheorem}   The inverse map $h^{-1}$ of the  diffeomorphism   $ h: \cH_+ \to \cH'_+$
 (resp. $h: \cH_- \to \cH'_+$) is
given by
\beq  \label{the inverse} 
\begin{split}
& h^{-1}(X^\flat) =    \frac{1}{\sqrt{(d'  + \gd')(X^\flat)}} (h'+  \gh')(X^\flat)\\
& \hskip 2 cm  \bigg(\ \ \text{resp.}\ \   h^{-1}(X^\flat) \= -   \frac{1}{\sqrt{(d' + \gd')(X^\flat)}}  (h'+\gh' )(X^\flat)\ \bigg).
\end{split}
\eeq
\end{theo}
\par \medskip
\begin{rem}  If the special Vinberg cone  $\cV$ is symmetric (see Definition \ref{self-adjoint}), the rational map $ \gd'$ vanishes and $d' + \gd' = d'$ is a cubic polynomial. In this case,  the relation \eqref{formula}  corresponds
to  the  {\it adjoint identity} of its canonically associated cubic Jordan algebra  (see e.g. \cite[\S 3.8]{MC}).
Denoting by   $d_{abc}$ and $d'{}^{d e f}$ the   tensorial components   of $d$ and $d'$, respectively,  in a system of
standard coordinates $(y^a)$ for $\cH$, it    takes the  polarised form  (up to a normalisation)  found in  supergravity literature (see e.g. \cite{GST, CV, BMR} and references therein)
\beq\label{freudenthal}   d'{}^{a b c} d_{a (df|} d_{b  |g h)}   = \d^c_{(d} d_{fgh)} \ .\eeq
\end{rem}
\par
\medskip
\section{Projective-special K\"ahler manifolds and very  special
 cones}
 \label{projective-bis}
The scalar manifolds (i.e. the target spaces of the maps  representing the scalar field sector of the Maxwell-Einstein 4-dimensional supergravity theories
considered in this paper) are K\"ahler manifolds of a  particular kind, the so-called {\it projective-special K\"ahler manifolds}.
For  reader's convenience,  we    briefly review  some properties  of these manifolds and  of their fundamental relations   with the  special Vinberg cones (for further information, see   \cite{Ce,Co, CDJL}; see also the extensive discussion in   \cite{LS} and references therein).
\par
\smallskip
\subsection{Conical  scalar manifolds, projective  scalar manifolds and special Vinberg cones} \label{projective-special}
 A {\it conical affine special K\"ahler }  (for short, {\it conical scalar}) {\it manifold}  is a  K\"ahler manifold $(\cM, J, g_\cM)$ equipped with  a  flat torsion free connection $\n$  and a homothetic vector  field $\xi$ such that:
 \begin{itemize}
\item[(i)] $\n (g_\cM \circ J) =0$;
 \item[(ii)] $(\n_X J)Y = (\n_YJ)X$;
 \item[(iii)] $\n \xi = D \xi = \Id$ where $D$ is  the Levi-Civita connection of $(\cM, g_\cM)$;
 \item[(iv)] the metric $g_\cM$ is positively defined on   $\cD \=   \Span(\xi, J \xi)$ and it is negatively defined on   $\cD^\perp$;
  \item[(v)] the commuting holomorphic vector fields $\xi, J\xi$ are complete and define a free holomorphic   $\bC^*$-action
  $$ (\r e^{i \q}, x) \longmapsto t e^{i \q}{\cdot}x \= (e^{t \xi} \circ e^{\q J \xi})(x)\ ,$$
 in which    $\{x \mapsto e^{i \q} {\cdot} x = e^{\q J \xi}(x)\}$  is a one-parameter isometry group.
 \end{itemize}
In this case     the hypersurface
   $$ \wt \cS := \{  x \in \cM, \ \  g_\cM(\x, \xi )|_\cM=1 \}\ \subset \cM$$
    is
    $S^{1} = \{e^{i \theta}\}$-invariant, with an induced Lorentzian   metric $ g|_{T\wt \cS}$ of signature $(1, 2n)$ and   a sub-Riemannian structure  $(\langle J \xi \rangle^\perp, - g_{\wt \cS}\big|_{\langle \langle J \xi \rangle^\perp} )$ of rank $2n$, which are both preserved by the $S^1$-action.   By $S^1$-invariance, the sub-Riemannian structure $(\langle J \xi \rangle^\perp, - g_{\wt \cS}\big|_{\langle \langle J \xi \rangle^\perp} )$ projects onto a K\"ahler structure  $(J, g_{\cS})$  on the   quotient  $\cS =  \wt \cS/ S^1 \simeq \cM/\bC^*$ (\cite{ACM}).\par
\smallskip
   The  K\"ahler manifolds $(\cS, J, g_{\cS})$ determined  in this way  are called  {\it projective special K\"ahler manifolds} or,  motivated by
   physics terminology,   {\it projective scalar manifolds}. \par
\par
\smallskip
\subsection{Description of  projective scalar manifolds in terms of  conical  holomorphic coordinates and prepotentials}\label{local}
  Let $((\cM, J, g_\cM),  \nabla, \xi)$ be a   conical scalar  manifold (= conical affine special K\"ahler manifold)   of  $\dim_\bR \cM = 2n+2$ and  with  associated projective scalar $2n$-manifold $(\cS, J, g_{\cS})$.   Around any point  of $\cM$, there exists a neighbourhood   $\cU \subset  \cM$, on which there exist  a distinguished system of holomorphic coordinates $ X = (X^I),  \, I = 0, 1, \cdots , n$, called {\it conical special coordinates},   and  a  homogeneous  of degree $2$ holomorphic  function    $F: \cU \to \bC$,  called {\it prepotential},   which  locally determines all data of   $\cM$ as  follows.
   \begin{itemize}[leftmargin = 15pt]
 \item[(a)] the complex structure $J$  is given by   the  multiplication by $i = \sqrt{-1}$ on the  holomorphic vector fields $\frac{\p}{\p X^I}$;
 \item[(b)]   $g_\cM$  is the metric given by
  \beq \label{metric-prep} g_\cM = m_{IJ}dX^I \otimes d\overline X^J \  \text{with}\ \  m_{IJ}(X, \overline X) \=  2   \Im(F_{IJ}) = 2  \Im\left( \frac{\p^2  F}{\p X^I \p X^J} \right)\bigg|_{(X, \overline X)}\eeq
  and has  the real function  $\f = - \log(g_{\cM}(\xi, \xi)|_\cU)$ as K\"ahler potential;
   \item[(c)] $\xi$  is   the vector field $  \xi = X^I \frac{\p}{\p X^I} + \overline{X}^I \frac{\p}{\p {\overline X}^I}$.

   \item[(d)]  using the notation   $x^I \= \Re(X^I)$, $y^J\= \Im(X^J)$ and $y_{J}\= \Re(F_J)$, the connection $\n$ is uniquely determined by the conditions  that $x^I$ and $y_J$ are $\n$-flat, i.e. $\n {\frac{\p}{\p x^I}}= \n {\frac{\p}{\p y_{J}}} = 0$.  Note that, since   $y_{J} \neq y^J$,  in general   the   vector fields $\frac{\p}{\p X^I}  = \frac{\p}{\p x^I} - i \frac{\p}{\p y^I}$ are   {\it not  $\n$-flat}.
   \end{itemize}
\par\smallskip
The  $\bC^*$-invariant  functions  $z^a = \frac{X^a}{X^0}$, $1 \leq a \leq n$,  are holomorphic  coordinates on the corresponding open subset $\wt \cU \subset \cU/\bC^*$  in    the  projective-special K\"ahler manifold $(\cS = \cM/\bC^*, J, g_\cS)$ and   any  $\bC^*$-invariant map  $f: \cU \to \bR$    corresponds to a unique function $f_{\wt \cU}: \wt \cU \subset \cS\to \bR$ given by
  $$f_{\wt \cU}|_{z = [X]} = f (1, X^1, \cdots, X^n)$$
 (in other words, in physics jargon,  the K\"ahler gauge freedom is fixed such that $X^0 = 1$).
In the open subset $\wt \cU \subset \cS$,  the metric $g_{\cS}$ has the   K\"ahler potential
 $$\cK|_{z= [1: X^1:  \ldots: X^n]_{\bC^*}}\= -   \log \sum_{I,J =0}^n X^I m_{IJ}(X, \bar X) \bar{X}^J\ .$$
\par
\smallskip
 \subsection{The supergravity r-map} \label{veryspecial}
 Let $W = \bR^{n+1}$ be an $n+1$-dimensional vector space. A  homogeneous  cubic polynomial $d: W \to \bR$ is called  {\it hyperbolic}  if
the Hessian $\p^2 d_x $ has  signature $(1,n)$ at some  $x \in W$.
A hypersurface  $\cT \subset W$ is called  {\it very special real}  if
 there exists  a  homogeneous  cubic  polynomial
$d$ such that  $\cT \subset \cT_d = \{ d =1\} $  and the tensor field $g_{\cT} : =- \p^2 d |_{T\cT}$ is  a Riemannian metric on $\cT$. Note that if this occurs, then  $d$ is  hyperbolic.  \par
 The  {\it supergravity r-map} is the correspondence which   associates with  each  very special real manifold   $(\cT, g_{\cT})$  the  projective  scalar manifold
 $(\cS_\cT, J, g_{\cS_\cT})$  where: (i)  $\cS_\cT$ is the open submanifold
 $$   \cS_\cT := \bR^n + i\cV  \subset \bC^n  ,\qquad \cV = \bR_+ {\cdot} \cT\ ,$$
 called {\it Siegel domain},  and
(ii)  $J$ is the  standard complex structure of $\bC^n$  and (iii) $g_{\cS_\cT}$ is the  K\"ahler metric, given in terms of the standard coordinates $(z^a = x^ a + i y^a)$ on $\bC^n$  by
\beq g_{\cS_\cT} = \p_{z^a}\p_{z^b} \cK(z, \overline z) d z^a  \otimes d \overline z^b\ ,\qquad \cK \= -  \log 8 d(y)  \ .\eeq
The triple  $(\cS_\cT, J, g_{\cS_\cT})$ is  the  projectivisation of the
  conical  scalar  manifold
  $$\cM_\cT \=  \{ \ X = X^0 \cdot (1,z^1, \ldots, z^n) \ , \ X^0 \in \bC^*, \, z  \in \cS = \bR^n + i \cV\   \} \subset \bC^{n+1}$$
  with   K\"ahler metric, homothetic vector field and connection $\n$, which are uniquely determined by  the     coordinates
  $$(X^I) = (X^0, X^1 = X^0 {\cdot} z^1, \ldots, X^n = X^0 {\cdot}  z^n)$$
   of $\bC^{n+1}$ and the prepotential
  $$   F : \cM_\cT \longrightarrow  \bC\ , \qquad    F(X^I)\= \frac{d(X^1, X^2, \ldots , X^n)}{X^0}
$$
as   in (a) --(d) of \S \ref{local}.
\par
The  projective scalar manifold   $(\cS_\cT = \cM_\cT/\bC^*, J, g_{\cS_\cT})$   is    {\it the image of the special cubic  $(\cT, g_\cT)$ under the $r$-map}. Note that, up to local equivalences,   the $r$-map is essentially surjective, namely
\par
\begin{theo}[{\cite[Prop. 1.6 \& 1.10]{Co}}]  \label{theo31} Any $2n$-dimensional projective scalar  manifold $(\cS, J, g_\cS)$ is locally isometrically biholomorphic to the image under the $r$-map $ (\cS_\cT,J, g_{\cS_\cT})$   of a special cubic $(\cT, g_\cT)$ in $W= \bR^n$.
\end{theo}
\par
\medskip
If the special cubic cone $\cV = \bR_+ {\cdot} \cT$ of  a special cubic $(\cT, g_\cT) \subset \bR^n$ is convex and homogeneous, the special cubic manifold $(\cT, g_\cT)$ is called {\it Vinberg cubic}.
 \begin{theo}[\cite{WP1, Co}]  \label{theo32}  Let $(\cT, g_\cT)$ be a special cubic manifold  associated with an irreducible  polynomial  $d$ (that is,  not decomposable as the  product  of  two lower degree polynomials).  The  manifold $\cT$  is a Vinberg cubic   if and only if the corresponding  cone $\cV = \bR_+{\cdot} \cT$ is a   special Vinberg cone and $\cT = \{  d = 1\} \cap \cV$.
 \end{theo}
\par
\medskip
\section{BPS black holes in $\cN = 2$ $4D$ supergravity}
\subsection{Scalar fields,  vector fields and central charges in an $\cN = 2$ $4D$ supergravity theory }
\label{section31}
 As we mentioned in the introduction, our discussion is focused on the BPS black holes in theories of {\it ungauged $\cN = 2$ supergravity
 on   a  $4$-dimensional Lorentzian space-time $M$}.   For  comprehensive overviews of the contents of  such supergravity   theories we  refer  to  the foundational papers  \cite{WLPSP, WP, CKPDFWG} and to  the vast  subsequent literature  on the topic (see e.g.\cite{ABCDFFM, LS}  and references therein). For our  purposes we   need   to know just a  few   basic features, which we now  recall. \par
  For what concerns the  physical contents of  such supergravity theories, we are  interested only in the
    gravitational field (i.e. a Lorentzian metric $g$ on the space-time $M$)  and in  the bosonic sector of  the  vector multiplets, namely:
 \begin{itemize}[itemsep=5pt, leftmargin=15pt]
\item the   {\it scalar fields}, which are mathematically represented by the coordinate components of  a  map $z: M \to \cS$   into an appropriate  target complex $n$-manifold  $\cS$, called {\it scalar manifold};
\item the   {\it vector fields  of $n+1$ abelian gauge fields};  up to  gauge transformations these fields are given by  the  components   $A_\mu^I$, $0 \leq I  \le  n$,
 of  the potential of a connection  on a  principal bundle   $\pi: P \to M$ of the abelian  group
$T^{n+1} = \underset{n+1\ \text{times}}{\underbrace{S^1 \times \ldots \times S^1}}$.
 \end{itemize}
 \medskip
  \label{scalarsect}
 \label{subsubsection-scalar}
 {\it The  scalar manifold   $\cS$    is always  assumed to be a projective scalar manifold $(\cS = \cS_\cT, J, g_{\cS})$ of complex dimension $n$ which is image of  a  special cubic  $(\cT, g_\cT)$ under the  $r$-map}, as  defined in
 \S \ref{projective-special}.  Such a scalar manifold $\cS = \cS_\cT$ is therefore  a complex manifold of the form  $\cS = \bR^n + i \cV$ where  $\cV = \bR_+ {\cdot} \cT \subset \{ d > 0\}$   is a special cubic cone  associated with  $\cT \subset \{ d = 1\} \subset \bR^n$. By Theorem \ref{theo32},  if  $ d $ is irreducible and $\cV$ is convex and homogeneous, then  $\cV$ is a  special Vinberg cone  according to the definition in  \S \ref{specialVinbergcones}.
 \par
{\it  The field strengths of the abelian gauge fields  (= the electromagnetic fields)} of the theory  are geometrically given  by  the $n+1$ components   of the curvature $2$-form
 $$\bF = \left(\bF^0_{\mu \nu} d y^\mu \wedge dy^\nu, \ldots, \bF^n_{\mu \nu} d y^\mu \wedge dy^\nu\right)  \ ,\qquad \text{with}\quad  \bF^I_{\mu \nu}\= \frac{1}{2}\left(\frac{\p A^I_{\mu}}{\p y^\nu} - \frac{\p A^I_{\nu}}{\p y^\mu}\right)\ ,$$
 where   $(y^{\mu})$  are  coordinates for the  space-time $M$.
 \par
\smallskip
In addition to the scalar and the abelian gauge fields,  there is another  important function  which has  to   be considered:  {\it the central charge} $Z$.  It  is a  $\bC$-valued field $Z: M \to \bC$ on the space-time which  is   determined
  by the  physical  fields of the theory and  satisfies a  continuity equation.   \par
  \smallskip
   The notion of the central charge  was first  introduced in   \cite{WO, SW} for generic solutions of   supersymmetric theories  and  gives important  information on the physical properties of the solutions of the  field equations. For an extensive discussion,  we refer the interested  reader to the original papers. For our purposes, we   need  to recall just  a  couple of      facts concerning   its relation with the entropy of  the black holes. We  briefly  review them  in   \S \ref{section3.3}.
   \par
   \smallskip
\subsection{Static and spherically symmetric  black holes and their  electro-magnetic charges}
\label{section3}
 \label{static}
From now on, we  assume to be working in a  fixed supergravity  theory on a $4$-dimensional space time of the kind described in \S \ref{section31}.  Moreover, by  ``black hole''   we constantly understand a   {\it static},
{\it spherically symmetric and asymptotically flat gravity field}
 {\it that solves the equations of the theory and  has  the singularity of an isolated black hole}.  In mathematical terms  this means    that the  metric $g$ and the bosonic fields  associated  with the   black hole satisfy  the following conditions.
 \begin{itemize}[leftmargin = 20pt]
 \item[(i)]    The Lorentzian manifold $(M, g)$ admits  a set of globally defined  coordinates $(t, r, \q, \f)$, where:
 \begin{itemize}
 \item[(a)]  the vector field $\frac{\p}{\p t}$ is  time-like,
 \item[(b)]  each  level set   $M_{t = t_o} = \{t = t_o \}$ is    space-like  and diffeomorphic to $\bR^3$,
 \item[(c)]
  $(r, \q, \f)$   are    spherical coordinates for    each submanifold  $M_{t = t_o}   \simeq \bR^3$.
  \end{itemize}
 \item[(ii)]   The metric has  the  form
 \beq  \label{extremality-1}
g=-e^{2U(r)}dt^{2}+e^{-2U(r)}\left( dr^{2}+r^{2}\left( d\theta ^{2}+\sin
^{2}\theta d\varphi ^{2}\right) \right)
 \eeq
   such that $g$ tends to the flat metric for $r\rightarrow \infty $, and
\begin{equation}
e^{-2U(r)}\cong _{r\rightarrow 0^{+}}\frac{C^{2}}{r^{2}},\qquad\text{for some constant} \ C\in\bR,
\end{equation}%
so that there is an event horizon at $r=0$ (see for instance \cite{ADFT, FHM}). The geometry in the near-horizon limit $%
r\rightarrow 0^{+}$ is the Bertotti-Robinson geometry $AdS_{2}\times S^{2}$
\cite{Bertotti, Robinson}.  The area of the $S^{2}$ which equals the area of
the unique event horizon reads%
\begin{equation}
A_H=4\pi C^{2}.\label{area}
\end{equation}
   \item[(iii)] The scalar  fields and the electromagnetic fields   are $t$-independent and   invariant under the standard $\SO_3$-action on
   each  space-like submanifold  $M_{t = t_o}   \simeq \bR^3$.
        \end{itemize}
   By saying that \textquotedblleft there is a black hole horizon at $r=0$\textquotedblright\ we    mean that the value $r= 0$ determines   the  boundary of an $SO_{3}$-invariant and time independent region of the space-like manifold $M_{t=t_{o}}\simeq \mathbb{R}^{3}$, from which no light ray might escape. 
   \par
\medskip
  Given a black hole, for any   fixed  values    $t_o$ and $r_o > > 0$,   we may consider the  double integrals
\begin{eqnarray}
p^{I} &&:=\frac{1}{4\pi }\underset{\mathbf{S}=\left\{ t=t_{o},r=r_{o}\right\} }{\int}%
\hskip -0.5 cm \mathbb{F}^{I}  \nonumber \\
&&=\frac{1}{4\pi }\underset{\vartheta \in \lbrack 0,\pi ], \varphi \in \lbrack 0,2\pi ]}{\iint} \hskip -0.5 cm \mathbb{F}_{\vartheta \varphi
}^{I}d\vartheta \wedge d\varphi,  \qquad ~\text{with~}~\mathbb{F}_{\vartheta \varphi
}^{I}:=\left. \mathbb{F}^{I}\right\vert _{\mathbf{S}}\left( \frac{\partial
}{\partial \vartheta },\frac{\partial }{\partial \varphi }\right) , \\
q_{I} &&:=\frac{1}{4\pi }\underset{\vartheta \in \lbrack 0,\pi ], \varphi \in \lbrack 0,2\pi ]}{\iint} \hskip -0.5 cm \mathbb{G}_{I|\vartheta
\varphi }d\vartheta \wedge d\varphi, \hskip 0.35 cm ~\text{with~}~\mathbb{G}_{I|\vartheta
\varphi }:=\left( \left. \star \mathbb{F}^{I}\right\vert _{\mathbf{S}%
}\right) \left( \frac{\partial }{\partial \vartheta },\frac{\partial }{%
\partial \varphi }\right) ,  \nonumber \\
&&\label{charges}
\end{eqnarray}
 where $\bG$  is a tensor field which is uniquely determined by the electromagnetic field $\bF$ by means of an appropriate generalised Hodge star operator. This operator   depends in a non-trivial way on the scalar fields,   but for our purposes we do not need to know its explicit expression.
By    (iii) and  Gauss' Theorem, these integrals  are  independent on  the choice of   $t_o$ and   $r_o$ and  can be considered as   the    sources (= ``charges'')  of   the  $n+1$  electromagnetic fields $\bF^I \= \bF^I_{\mu \nu} d y^\mu \wedge d y^\nu$ at  large distances from the black hole. Indeed the $2$-forms $\bF^I$
 behave precisely as if they were classical electromagnetic  fields,   generated by  corresponding   magnetic monopoles and/or  electric charges, all of them  located at  $r = 0$. The  values  $p^I$,  $q_I$   are  called the   {\it  magnetic charges} and  the  {\it electric  charges}   of  the   black hole, respectively.  If there are two  charges $p^I$, $q_I$, which are   both non trivial,
the corresponding electromagnetic field  $\bF^I$ behaves as if it were   generated by  a dyon and the black hole is    called   {\it dyonic}.
\par
\smallskip
\subsection{BPS black holes,  their entropy and the ``inverse relation''  map}
\label{section3.3}
The first of the two properties of the central charge, which  we need to recall,  is that  {\it  for each  black hole satisfying (i), (ii),  (iii),   the function  $Z: M \to \bC$   depends   just on  the coordinate $r$  and,  for    $r\geq  0$,
 the   absolute value $ |Z(r)|$ is bounded from above by  the mass of the black hole}  {\rm (\footnote{By  {\it mass}  of a  black hole  we mean
its  {\it Bertotti-Robinson mass}  \cite{Bertotti, Robinson}.}) }
 \beq \label{inequa} m \geq |Z(r)|\ .\eeq
 The black holes  for which the inequality  \eqref{inequa}  is   ``saturated''   (i.e. such that $m=\left\vert Z(r_{o})\right\vert=: |Z_{o}|$  for some $r_o \geq 0$) are called  {\it BPS} (\footnote{The acronym     {\it BPS}
 stands for Bogomol'ny\u\i, Prasad and Sommerfield, who were  the first  that determined  explicit  solutions of  supersymmetric equations,  where an inequality  of the form     \eqref{inequa}  appeared as   saturated  \cite{PS, Bo}.}). \par
 \medskip
On the other hand, by the Bekenstein-Hawking entropy-area formula (\cite{Ha, Be}),   the  thermodynamical entropy $S$ of a black hole
is completely determined by the area $A_H$ of the   (unique and time independent) event horizon located at  $r = 0$. More precisely, in natural units
\beq S = \frac{A_H}{4} . \eeq
This relation, together with Eq. (\ref{area}) and with the fact that, for an
extremal BPS black hole (see for instance \cite{ADFT, FHM})%
\begin{equation}
C^{2}=|Z_{o}|^{2},%
\label{area-2}
\end{equation}%
implies that for such a black hole the following relation \ between the
entropy and the central charge holds :%
\begin{equation}
S=\pi |Z_{o}|^{2}.\label{theformula}
\end{equation}
\par
\smallskip
 The second important  property of the central charge that we need to recall is  a crucial   phenomenon   of the    BPS black holes.   Consider one such black hole,   with magnetic and electric charges $p^I$,  $q_J$ and mass $m \neq 0$.  We recall that the scalar fields associated with such a (static and spherically symmetric)  black hole are given by a map, depending only  on the radius $r$,
 $$z = z(r): M \to \cS_\cT = \bR^n + i \cV = \bR^n + i \bR_+ \cdot \cT$$
into the projective scalar manifold $\cS_\cT =   \cM/\bC^*$,   determined from a special cubic cone of $\cT \subset \{ d  = 1\} \subset \bR^n$  through the $r$-map,
 which  is the projectivisation of a corresponding conical scalar manifold   $\cM \subset \bC^{n+1}$.  The map $z = z(r)$  is canonically associated  with  the   map  into the conical scalar manifold $\cM$
 $$X = X(r):  M \to \cM \subset \bC^{n+1}\ ,\qquad X(r) \= (1, z^1(r), \ldots, z^n(r))\ .$$
 We also recall that  the potential of the K\"ahler metric $g_{\cS_\cT}$ of the scalar manifold $\cS_\cT$  is the function $$\cK(z, \bar z) = - \log(8 d(\Im(z))$$
  and that the prepotential $F$, which characterises the conical scalar manifold $\cM \subset \bC^{n+1}$,  is the holomorphic function
  $$F(X) =  \frac{  d (X^1, X^2, \ldots X^n)}{X^0} = \frac{d_{abc} X^a X^b X^c}{X^0}\ .$$
 Let   now  $r_o \geq 0$ be  the smallest radius such that  $m = |Z_{o}|$  and   denote by
 $$Z_o: = Z(r_o) \in \bC\ ,
 \qquad z_o \= z(r_o) \in  \cS\ , \qquad X_o = (1, z^1(r_o), \ldots, z^n(r_o))\in \cM$$
     the values at $r_o$  of  the central charge $Z(r)$, of the scalar fields map $z(r)$ and   of the  canonical  associated lifted map  $X(r) = (1, z(r))$, respectively.
 The following crucial properties  hold:
  \begin{itemize}
  \item[(a)]  {\it   $r_{o}=0$, implying that $|Z_{o}|:=\lim_{r\rightarrow 0^{+}}|Z(r)|$};
  \item[(b)] denoting  $ F_I \= \frac{\p F}{\p X^I}$
\beq  \label{fundamental-bis}
\begin{split}
&p^K= - e^{\frac{\cK(z, \bar z)}{2}} 2\Im(\overline Z_o X_o^K) = - \frac{2\Im(\overline Z_o X_o^K)}{\sqrt{8 d (\Im(z))}}  \ ,\\
 &q_L=   - e^{\frac{\cK(z, \bar z)}{2}} 2\Im(\overline Z_o F_L|_{(X^J_o)}) = - \frac{ 2\Im(\overline Z_o  F_L|_{(X^J_o)})  }{\sqrt{8 d (\Im(z))}}\ .
\end{split}
\eeq
\end{itemize}
{\it  The  equalities \eqref{fundamental-bis}   completely determine the magnetic and electric charges  in terms
 of the scalar fields and the central charge of the BPS black hole} and they are consequences of  the celebrated  theory of the  {\it attractor mechanism}   \cite{FS, FK}.
We call them  {\it BPS relations}.
\par
We derive our main results  from   \eqref{theformula} and \eqref{fundamental-bis}. In fact,   in the next section,  we will show that  the problem of
inverting    the relations \eqref{fundamental-bis}   boils down to determining  the inverse map to the quadratic map
$$y \mapsto h (y) = (h_a(y) \= d_{a b c} y^b y^c)\ .$$
  On the other hand, in Theorem \ref{thetheorem}
 we determined the explicit expression of such inverse map in case $d$ is   an invariant cubic polynomial. Combining these results,   we are able to  obtain the
explicit expression for the modulus  $|Z_{o}|$ of the  central charge of a BPS black hole  --  and  thus, due to  \eqref{theformula},   for its entropy $S$ ---     in terms
 of its electric and magnetic charges only, provided  that the  scalar manifold $\cS = \bR^n + i \cV$ is  homogeneous.  The  formula we determine is valid for  {\it any choice of the  homogeneous scalar manifold}  and extends the  previously known  expression,    determined   only   for the cases  in which   $\cS_\cT$ is a  {\it symmetric manifold}.   \par
 \par
\medskip
\section{Recovering   the central charge and  the scalar fields
 from the electric and magnetic charges  of  a BPS black hole}
\subsection{The  BPS relations as  a map}
\label{section32bis}
In this section, we  discuss some aspects of the map,  which determines the BPS relations between central charge and scalar fields on one side and the magnetic and electric charges on the other side.
 In order to make fully  clear  that here and in the next two subsections  we   address purely   mathematical properties of this map -- whose arguments
and values  are  tuples of just (real or complex) numbers,  {\it  not  quantities with a  prescribed physical meaning} -- we adopt the following notational conventions.
\begin{itemize}[leftmargin = 15pt, itemsep = 5 pt]
\item An index  running from $0$ to $n$ (resp. from  $1$ to $n$) is  always denoted by a  capital  letter  as  $I$, $J$, $K$, etc.  (resp.   small  letter as  $a$, $b$, $c$, etc.).
\item  When no ambiguity may occur,   $n$-tuples  as $(z^a)$, $(\gp^a)$,  etc.,  are denoted   with no index,  i.e. by  $z$,  $\gp$, etc.
\item The standard  pairing  between a  $1$-form   $w = (w_a) \in \bR^n{}' \=  \Hom(\bR^n, \bR)$  and a vector $y = (y^a) \in \bR^n$  is  denoted by  $\langle w, y\rangle \= w_a y^a$.
\item The standard {\it complex} coordinates of $\bC^{n}$ and the standard
{\it real} coordinates of $\bR^{2n+2}$ are indicated  by $z = (z^a)$
 and by $ (\gp^0, \gp^a, \gq_0, \gq_a) = (\gp^0, \gp, \gq_0, \gq)$,
respectively
\item We  denote by $h(y) = \big(h_a(y) \= d_{a b c} y^b y^c\big) $
  the {\it quadratic  map}  determined by the cubic polynomial $d (y) = d_{abc} y^a y^b y^c$ of the scalar manifold $\cS_\cT = \bR^n + i \cV$. We use
the  same symbols  for their natural extensions
  $d : \bC^n \to \bC$ and $h: \bC^n \to \bC^{n}{}'$,   defined  by $d (z) \= d_{abc} z^a z^b z^c$ and $(h_a(z)) \= (d_{abc} z^b z^c)$, respectively.
\end{itemize}
\par
\medskip
From now on, we    focus on  the {\it BPS map}  given in  \eqref{algebraic},   namely the map
\begin{multline}\label{functionf}  \gf: \bC^* \times (\bR^n + i \cV)  \subset \bC^{n+1}\longrightarrow \bR^{2n+2}\ ,\\
\gf(Z, z^a, \overline Z, \overline z^a)  =  (\gp^K(Z, z^a, \overline Z, \overline z^a), \gq_L(Z, z^a, \overline Z, \overline z^a))\ ,
\end{multline}
defined by
\beq \label{inverserel}
\begin{split}
&\gp^0(Z, z, \overline Z, \overline z)\=    \frac{\Im Z}{\sqrt{2d(y)}}\ ,\\
&\gp^a(Z, z, \overline Z, \overline z)\= - \frac{\Im(\overline Z z^a)}{\sqrt{2d(y)}}\ ,\\
&\gq_0(Z, z, \overline Z, \overline z)\=  -  \frac{\Im\big(\overline Z d_{abc} z^a z^b z^c\big)}{\sqrt{2d(y)}}\ ,\\
&\gq_a(Z, z, \overline Z, \overline z) \= -   \frac{3 \Im\big(\overline Z d_{a bc} z^b z^c\big)} {\sqrt{2d(y)}}\ , \hskip 2 cm \text{where}\ \ y \= \Im(z)\ .
\end{split}
\eeq
The BPS relations  \eqref{fundamental-bis} tell that the value $\gf(Z_o, z^1_o, \ldots, z^n_o)$ of $\gf$ on  the central charge and the scalar fields map  at the horizon $r_o =0$ of a BPS black hole
gives  the magnetic and electric charges of  the  black hole
$$(p^K, q_L) = \big(\gp^K(Z_o, z_o, \overline Z_o, \overline z_o), \gq_L(Z_o, z_o, \overline Z_o, \overline z_o)\big)\ .$$
In what follows, we decompose  the  standard coordinate of $\bC^*$ as $Z = t e^{i \q}$,  with   $t = |Z| > 0$ and $\q \in \bR \!\!{\mod 2 \pi}$, and we use the short-hand notation  $\gc \=  \frac{1}{\sqrt{2d(y)}}$.
In this way, using the above notational convention,  the  \eqref{inverserel}  take the  form
\begin{align}
\label{1-1} & \gp^0 =   \gc t \sin \q\ ,\\
\label{2-1}  & \gp =   - \gc t \Im(e^{-i \q}z)\ ,\\
\label{3-1}   & \gq_0 =    \gc t  \Im\big( e^{- i \q}   \left(d(z)\right)\big)  \ ,\\
\label{4-1}  & \gq =   - 3 \gc t   \Im\big(   e^{- i  \q}h(z)\big)
\ .
\end{align}
We now consider  the purely mathematical questions  of finding  the domains  $\cU \subset \bC^* \times  \cS $ on which the BPS map $\gf$  is locally invertible and, on such domains,  determining
the explicit expressions for the (local) inverses of this map.
For this purpose,  it is convenient to split the domain $\bC^* \times \cS$ of the BPS map  into the union of the following four  disjoint subsets:
\begin{itemize}[leftmargin = 20pt]
\item[--]    the  $(2n+1)$-dimensional  hypersurfaces
$$\cC_+ \= \{\q  = 0\}\times \cS \ ,\qquad \cC_- \= \{\q = \pi\}\times  \cS\  ,$$
\item[--]  the  $(2n+2)$-dimensional open subsets
$$\cA_+ \= \{\q \in (0, \pi)\} \times \cS\ ,\qquad \cA_- \= \{\q \in (-\pi, 0)\} \times \cS\ .$$
\end{itemize}
We  discuss  the behaviour of the  restrictions $\gf|_{\cC_\pm}$ and $\gf|_{\cA_\pm}$ separately.
\par
\smallskip
\subsection{The  maps $ \gf|_{\cC_\pm}$ take  values  into  $\{ \gp^0 = 0\}$  and are globally  invertible}
Due to \eqref{1-1},  the map $ \gf$ sends both    hypersurfaces   $\cC_\pm $  into the
hypersurface  $ \{ \gp^0 = 0\}$ of  $\bR^{2n+2}$.  Moreover we have the following
\begin{prop}  Each map
 $$\gf|_{ \cC_{\pm}}:  \cC_\pm \to \gf(\cC_\pm) \subset   \{\ \gp^0 = 0\ \}$$
 is a  diffeomorphism onto its image. More precisely,  if     $(\gp^0  = 0, \gp ,  \gq_{0}, \gp)$ is a point of $\gf(\cC_{\pm})$, then
 there is a unique   $(t, z) \in \cC_{\pm}$ such that $(\gp^0 = 0, \gp,  \gq_{0}, \gp) = \gf(t,z) $. This point  is given by
 (\footnote{ Note that \eqref{3-1****}  gives  the well known    expression  $
S = \pi \sqrt{\frac{1}{3}d(\mathfrak{p})\left( \left\langle \gq,
\mathcal{D}\right\rangle +12\mathfrak{q}_{0}\right) }\label{entr-p0=0}
$   for the entropy  of a BPS black hole  with $\gp^0 = 0$ due to  Shmakova \cite{Sh}. We also  remark that, from \eqref{shma} and  the fact that   $d(\gp)<0$,  we  also have that   $z =\frac{1}{6} \cD(\gp,\gq) +i \frac{\gp}{2} \frac{S}{\pi}  \frac{1}{d(\gp)}$.
This  amends an error in \cite[Formula (27)]{Sh}})
 \begin{align}
 \label{shma} & z  =   \frac{1}{6}  \cD(\gp_, \gq)  - i \gp \sqrt{   \frac{ \langle \gq ,  \cD(\gp, \gq) \rangle  + 12  \gq_{0}}{ 12 d(\gp)}} \ ,\\
 & \label{3-1****} t = \left( \frac{  d(\gp) \big(\langle \gq ,  \cD(\gp, \gq) \rangle  + 12  \gq_{0}\big)}{12}\right)^{\frac{1}{4}}\ ,
 \end{align}
where $\cD^a(\gp, \gq) \= \cD^{ab}(\gp) \gq_{b}$ with   $ \left( \cD^{ab}(\gp)\right)$  inverse matrix of  $ \left(d_{a b c} \gp^c\right)$.
\end{prop}
\begin{pf}  We give the  proof   only for   $\gf|_{\cC_+ }$, the other being  similar. Given  $(\gp^0 = 0, \gp,  \gq_{ 0}, \gp) \in  \gf(\cC_+)$,  any pre-image $(t, z = x + i y) \in \cC_+$ satisfies
\begin{align}
\label{2-1*}  & \gp = -  \gc t  y\qquad \Longrightarrow\qquad y =  - \frac{\gp}{\gc t}\ , \\
\label{3-1*}   & \gq_{0} =      3   \langle h(x), \gc t  y\rangle -  \frac{1}{\gc^2 t^2} d(\gc t y)  \ ,\\
\label{4-1*}  & \gq =    - 6    d_{\cdot b c} x^b (\gc t y^c)  \ ,\qquad \text{with}\ \ \gc \=  \frac{1}{\sqrt{2d(y)}}
\end{align}
Replacing \eqref{2-1*} into \eqref{3-1*}  and  \eqref{4-1*}, we get
\beq
\label{3-1**}  \gc t =  \sqrt{  \frac{d(\gp)}{3   \langle h(x),  \gp\rangle +  \gq_{ 0} }}   \ ,\qquad \gq =    6    d_{\cdot bc} x^b  \gp^c    \ .
\eeq
 It follows that
\begin{multline}
\label{3-1***} x =   \frac{1}{6}  \cD^{\cdot b}(\gp) \gq_{ b}\ ,\quad  \gc t =   \sqrt{  \frac{ 12 d(\gp)}{ \langle \gq ,  \cD^{\cdot b}(\gp) \gq_{b} \rangle  + 12  \gq_{0}}}  \ ,\\
 y = -  \frac{\gp}{\gc t}   =  - \gp\sqrt{  \frac{ \langle \gq ,  \cD^{\cdot b}(\gp) \gq_{b} \rangle  + 12  \gq_{0}}{ 12 d(\gp)}}\ .
 \end{multline}
 and
$\gc^2 =  \frac{1}{ 2d(y)} =   12 \sqrt{\frac{ 3 d(\gp)}{     (\langle \gq ,  \cD^{\cdot b}(\gp_0) \gq_{b} \rangle  + 12  \gq_{0})^{3}}}$.
This  and $\eqref{3-1***}$  imply the claim.
\end{pf}
\par
\smallskip
\subsection{The  maps $\gf|_{ \cA_\pm} $ take values into  $\{ \gp^0 \neq 0\} $ and are  locally invertible}
The situation for  the restriction of $\gf$ on  the  sets $  \cA_\pm  \subset \cS \times \bC^* \setminus (\cC_+ \cup \cC_-) $ is  quite different from the previous. First of all, by   \eqref{1-1},
it is clear that  the images $\gf(\cA_\pm)$ are included   in the open subset $\{ \gp^0 \neq 0\} \subset \bR^{2n+2}$.  Moreover
 \begin{theo} \label{main1}  Each  map $\gf|_{ \cA_{\pm}}: \cA_\pm  \to \gf (\cA_{\pm}) \subset \{\gp^0 \neq 0 \}$
is a  local diffeomorphism and   the (locally defined) inverse map
 $$(\gp, \gq) \to (t e^\q, z)  $$
  is given by
\begin{align}
\label{5-1} &  z \= \frac{e^{i\vartheta }}{\mathfrak{p}^{0}} h^{-1} \left( h\left( \mathfrak{p
}\right) - \frac{1}{3}\mathfrak{p}^{0}\mathfrak{q}\right) +\frac{\gp}{\gp^0}  ,\\
\label{6-1}   &\q \= \pm \arccos\left( \frac{\gp^0\left( \gp^0\gq_0+\left\langle \gq,\gp\right\rangle \right) -2d\left(
\gp\right)}{2( \gp^0)^3  \Big\langle  h(\gp) -\frac{1}{3} \gp^0\gq, h^{-1}( h(\gp) -\frac{1}{3} \gp^0\gq) \Big\rangle}\right)\\
\label{7-1} & t \= \sqrt{2}  \sqrt{ \frac{\sin \vartheta}{\gp^0} \Big\langle  h(\gp) -\frac{1}{3} \gp^0\gq, h^{-1}(  h(\gp) -\frac{1}{3} \gp^0\gq) \Big\rangle}\ .
\end{align}
 More precisely, given    $(t_o e^{i \q_o}, z_o) \in \cA_\pm$ and its image   $(\gp^I_o, \gq_{oJ}) = \gf(t_o e^{i \q_o}, z_o) $,  there exist
 two neighbourhoods $\cU$,  $\cU'$ of  $(t_o e^{i \q_o}, z_o)$ and $(\gp^I_o, \gq_{oJ})$, respectively,
such that the restricted map $\gf|_\cU$ is a diffeomorphism from $\cU$ into $\cU'$ and its inverse $\gF = (\gf|_{\cU})^{-1}: \cU' \to \cU$ is given by   \eqref{5-1} -- \eqref{7-1}. \par
The neighbourhood $\cU' \subset \bR^{2n+2}$ of $(\gp^I_o, \gq_{oJ}) $ is  defined as follows.
 Consider  the vector $v_o$ and the $1$-form $\a_o$ defined by
 $$v_o \= \frac{ t_o y_o} {\sqrt{2d(y_o)}} \in \cV_d\subset  \bR^n \ , \qquad \a_o \=   3h( \gp_o) -\gp^0_o \gq_o \in \bR^{n*}\ .$$
  Then $\a_o$ belongs to  the cone $h(\cV_d)  \subset  h(\{ d > 0\})$ and
   there are  two  neighbourhoods $\cW_1 \subset \bR^n$  and $\cW^*_2 \subset \bR^{n*}$   of  $v_o$ and  $\a_o$, respectively,  such that  the restriction  $ h|_{\cW_1}: \cW_1 \to \cW_2^*$
  is a diffeomorphism between such two neighbourhoods. Then $\cU'$ is  the open set
 \beq \label{Uprime} \cU'\=  \{ (\gp^I, \gq_J) \ :\ \gp^0\neq 0\ , \
 \   3h( \gp) -\gp^0\gq \in \cW_2^*\  , \quad \operatorname{sign}(\gp^0) = \operatorname{sign}(\gp^0_o)\ \}\ .\eeq
 The corresponding neighbourhood $\cU \subset \bC^{n+1}$  of $(t_o e^{i \q_o}, z_o)$ is given by the images of $\cU'$ under the map \eqref{5-1} --- \eqref{7-1}.
 \end{theo}
 \begin{pf}
  (1)  Consider the change of    coordinates on $\{\Im(z) \neq 0\} \times \bC^n \subset \bC^{n+1}$
   $$(t e^{i \q}, z^a) \longmapsto (t e^{i \q}, \wt z^a  =   e^{-i \q}z^a)\ .$$
   In these new coordinates, the   components   \eqref{2-1} and \eqref{4-1} of  $\gf$ become  (here, $\wt x \= \Re(\wt z)$ and $\wt y \= \Im(\wt z)$)
\beq \label{4-1bisbis} \gp =  - \gc t \wt y\ ,\qquad
  \gq =   3 \gc t \Re(i e^{i \q}  h(\wt z))\ .
\eeq
 Using the first   of these two relations,  we may rewrite  the second  as
 \begin{multline}\label{325bis}
\gq =  3 \gc t \Re\left((i \cos\q - \sin \q) \left(h(\wt x)-  h(\wt y)  +2 i  d_{(\cdot) b c} \wt x^b \wt y^c\right)\right) = \\
=   3 \gc t \left( -\sin \q\, h(\wt x)  -  2 \cos \q d_{(\cdot) b c} \wt x^b \wt y^c \right) + 3 \gc t \sin \q\,h(\wt y) = \\
{=}  3  \gc t\sin \q\, \left(- h(\wt x) -2 \frac{\cos \q}{\sin \q\, } d_{(\cdot) b c} \wt x^b \wt y^c - \frac{\cos^2 \q}{\sin^2 \q} h(\wt y) \right)
+ 3 \gc t  \frac{\sin^2\q + \cos^2\q }{\sin \q} h(\wt y)  {=} \\
= - 3  \gc t\sin \q h \left(\wt x +  \frac{\cos \q}{\sin \q} \wt y \right)
+ \frac{3}{\gc t \sin\q} h(\gp) \ .
\end{multline}
On the other hand,  from \eqref{1-1} we may replace $\gp^0 = -\gc t \sin \q$ at all points. Since  $y = \Im(e^{i \q} \wt z) = \sin\q\, \wt x + \cos \q\,\wt y$, we have that \eqref{325bis} implies
\beq \label{326}
\begin{split}  \frac{ 3  \gc^2 t^2}{\gp^0}h(y)  & = \frac{3 \gc t }{\sin\q}h\left(\sin \q\, \big(\wt x +  \frac{ \cos \q}{\sin\q} \wt y\big) \right)   = \\
& =   3  \gc t\sin \q h \left(\wt x +  \frac{\cos \q}{\sin \q} \wt y \right)   =   \frac{3}{\gp_0}  h(\gp) - \gq\ .
\end{split} \eeq
This means that, if  $(\gp^I_o, \gq_{oJ}) = \gf(t_o e^{i \q_o}, z_o)$ for some  $ (t_o e^{i \q_o}, z_o) \in  \{\Im Z \neq0\}) \times \cS$, the  $1$-form $\a_o \= 3 h(\gp_o) - \gq _o \gp^0_o\in \bR^{n*}$
is actually equal to
$$ \a_o =   3  \gc^2_o t_o^2 h(y_o) =   h (v_o) \ , \qquad v_o \= \sqrt{3}\gc_o t_o y_o$$
that is $\a_o$ is  the  image of  the element  $v_o$ of the cone $\cV_d$  under the map $h$.
We now observe that, at each point $y \in\cV_d$, the Jacobian of the map $h$ is
$ Jh|_{y} = 2\left(d_{a b c} y^c\right)$. Being  $d$  associated with a  special   cubic, this matrix  is non-degenerate.  Hence, by  the Inverse Function Theorem,   there exists a  neighbourhood $\cW_1$ of $v_o \in \cV_d$  and neighbourhood $\cW^*_2$ of $\a_o$
  such  the restriction $- h|_{\cW_1}: \cW_1 \to \cW_2^* $ admits an inverse $h^{-1}: \cW^*_2 \to \cW_1$.\par
\medskip
 We now claim  that if $(\gp^I, \gq_J)$ is  in the open subset  defined in \eqref{Uprime},
 then
 there exists  at least one point  $ (t e^{i\q}, z = x + i y) \in\{ \Im(Z) \neq 0\}\times  \cS $ which is mapped onto  $(\gp^I, \gq_J)$ by $\gf$.  Indeed,  by \eqref{1-1}, \eqref{4-1bisbis}  and \eqref{326},    if   $ (t e^{i\q}, z = x + i y)$  is  mapped onto  $(\gp^I, \gq_J)$, then
\begin{align}
\label{323}& t =  \frac{\gp^0}{\gc \sin\q}\ ,\qquad  \wt y = -\frac{\gp}{\gc t} = - \frac{\gp  }{  \gp^0}\sin \q\ ,\\
\label{324}&  3 (\gp^0)^2 h \left(\wt x +  \frac{\cos \q}{\sin \q} \wt y \right)   =  3   h(\gp) -  \gp^0 \gq\ .
\end{align}
Using  \eqref{323} and  being $h$ is quadratic,  \eqref{324} can be transformed into
$$   h \left( \gp^0\wt x -    \cos \q \gp  \right)   =   h(\gp) - \frac{1}{3} \gp^0 \gq \ .$$
Using the   inverse map  $h^{-1}: \cW_2^*  \to \cW_1 $, this condition becomes  equivalent to
\beq  \label{325} \wt x    =   \frac{1}{\gp^0}  h^{-1}\left( h(\gp) -  \frac{1}{3}\gp^0 \gq \right) +  \cos \q \frac{ \gp}{\gp^0} \eeq
Since  $z = x + i y = e^{i \q} (\wt x + i \wt y)$,  from \eqref{323} and \eqref{325} we get
\beq\label{326bis}  z = e^{i \q} A +\frac{\gp}{\gp^0}\ \qquad   \text{with} \ \ A \= \frac{1}{\gp^0} h^{-1} \left(h\left( \gp \right) -\frac{1}{3} \gp^0\gq\right)\ . \eeq
Plugging  this and  $ t =  \frac{\gp^0}{\gc \sin\q}$ into  \eqref{3-1}  and using $h {\circ} h^{-1} = \Id|_{\cW_2^*}$, we obtain
\begin{multline}  \gq_0 =   -\gc t  \Re\bigg( i e^{- i \q} d\big(e^{i \q} A +   \frac{\gp}{\gp^0}\big)\bigg)  = \\
=  -\gc t  \Re\big( i e^{2 i \q}   d(A) )
- \frac{\gc t}{(\gp^0)^3}  \Re\big( i e^{- i \q} d (p))
- 3 \frac{ \gc t}{\gp^0}  \Re\big( i e^{i \q}  \langle h(A), \gp \rangle\big)   = \\
= 2  \gc t  \sin \q \cos \q\, d(A)
- \frac{\gc t \sin\q}{(\gp^0)^3}  d (p)
+ 3 \frac{ \gc t \sin \q}{\gp^0}  \langle h(A), \gp \rangle  = \\
=   2  \cos \q\, \gp^0 d(A)
- \frac{d (p)}{(\gp^0)^2}
+   \frac{3}{ (\gp^0)^2} \langle  h(\gp) -\frac{1}{3}  \gp^0 \gq, \gp \rangle = \\
=  2  \cos \q\, \gp^0 d(A)
+ 2 \frac{d (p)}{(\gp^0)^2}
-    \frac{1}{ \gp^0} \langle  \gq, \gp \rangle
\  .
\end{multline}
Solving this equation  with respect to $\cos \q$ we obtain
\beq \label{6-1bis}
 \cos \q =   \frac{\gq^0}{2 \gp^0 d(A)}
- \frac{ d (p)}{(\gp^0)^3 d(A)}
+ \frac{ 1} {2 (\gp^0)^2 d(A)}   \langle  \gq, \gp \rangle\ .
\eeq
and hence
\beq  \label{sin}
\sin\q {=} \pm \sqrt{1 - \cos^2 \q} {=}
\pm \sqrt{1 - \left(\frac{\gq^0}{2 \gp^0 d(A)}
- \frac{d (\gp)}{(\gp^0)^3 d(A)}
+ \frac{ 1} {2 (\gp^0)^2 d(A)}   \langle  \gq, \gp \rangle\right)^2},
\eeq
the sign being equal to  $+1$ in case $\q \in (0, \pi)$ (that is,   in case $\gp^0 > 0$)  and equal to  $-1$ otherwise. In both cases
$\frac{ \sin \q}{\gp^0} > 0$.
In order to conclude, we only need to recall that
\begin{multline} \label{7-1bis} t =\sqrt{ \frac{(\gp^0)^2}{\gc^2 \sin^2\q}}  = \sqrt{ \frac{2 (\gp^0)^2 d(y)}{ \sin^2 \q}}  =   \sqrt{ 2(\gp^0)^2 \sin \q  d(A)} =    \\
= \sqrt{2 \frac{\sin \q}{\gp^0}  d(\gp^0 A)} =  \sqrt{ 2 \frac{\sin \q}{\gp^0}  d(h^{-1} \left( h\big( \gp \right) -  \frac{1}{3} \gp^0\gq\big))}   = \\
=  \sqrt{2}  \sqrt{ \frac{\sin \q}{\gp^0} \left \langle h\left( \gp \right)- \frac{1}{3} \gp^0\gq,  h^{-1} \big( h\left( \gp \right) - \frac{1}{3} \gp^0\gq\big)\right\rangle} \end{multline}
\par
\smallskip
Finally,
we observe that,  since $\frac{ \sin \q}{\gp^0} > 0$, the equality \eqref{7-1bis}  makes sense only if
\beq \label{ineq} \left \langle h\left( \gp \right) - \frac{1}{3} \gp^0\gq,  h^{-1} \left( h( \gp ) - \frac{1}{3}\gp^0\gq\right)\right\rangle > 0 \ .\eeq
\hfill\end{pf}
\begin{remark} \label{theremark1}
 A  few simple algebraic  manipulations of the  \eqref{5-1} -- \eqref{7-1} lead to the  same   formulas   for the absolute value of the central charge (hence, also for  the entropy $S$)
 and for the  scalar fields at the horizon of  a BPS black hole   obtained  in \cite{Sh}  by  a different line of arguments.  Indeed,    given a point $(\gp^I, \gq_{J}) = \gf (t e^{i\q}, z) \in \gf(\cA_\pm)$, let   $\mathcal X$ and $\D = (\D_a)$  Shmakova's  vector and $1$-form defined by
 \beq  \label{434}
\mathcal X  \=   \sqrt{3} h^{-1} \left( h\left( \gp \right) - \frac{1}{3}\gp^0\gq\right) = \sqrt{3}\mathfrak{p}^{0}A \ ,\qquad
\Delta \= 3 \left(h(\gp)- \frac{1}{3}\gp^0\gq\right)\ .
\eeq
Then  \eqref{ineq} is equivalent to    $ \frac{1}{3^{\frac{3}{2}}} \langle \D,  \mathcal X\rangle > 0$.
Using this  and   \eqref{sin} we have that
\begin{multline} \label{7-1bisbis} t^2 {=}  \pm  \frac{2}{(3)^{\frac{3}{2}}}  \frac{1}{\gp^0}  \langle \D,  \mathcal X\rangle  \sqrt{1 - \left(\frac{ 3^{\frac{3}{2}} \gq^0 (\gp^0)^2}{2  d( \sqrt{3}\gp^0 A)}
{-}\frac{  3^{\frac{3}{2}}d (\gp)}{d( \sqrt{3} \gp^0A)}
+ \frac{ 3^{\frac{3}{2}}\gp^0} {2  d( \sqrt{3} \gp^0 A)}   \langle  \gq, \gp \rangle\right)^2} {=} \\
=   \pm  \frac{1 }{3 \gp^0} \sqrt{\frac{4}{3} \langle \mathcal X, \D \rangle^2 -  9 \bigg( \gp^0( \gq^0 \gp^0  +  \langle  \gq, \gp \rangle)
-  2d(\gp)
\bigg)^2} = \\
\overset{\text{the $\pm$ sign equals}\operatorname{sign}(\gp^0)}=    \frac{1 }{3 |\gp^0|} \sqrt{\frac{4}{3} \langle \mathcal X, \D \rangle^2 -  9 \bigg( \gp^0( \gq^0 \gp^0  +  \langle  \gq, \gp \rangle)
-  2d(\gp)
\bigg)^2}\ .
\end{multline}
This matches   \cite[Formula (12)]{Sh}.
Similar  straightforward algebraic manipulations of the \eqref{5-1} -- \eqref{7-1} lead to the expression
\beq
z =\left( \frac{1 }{\gp^{0}} \gp + \frac{3}{2}\frac{ \mathfrak{p}^{0}\left( \mathfrak{p}^{0}
\mathfrak{q}_{0}+\left\langle \mathfrak{q},\mathfrak{p}\right\rangle \right)
-2d\left( \mathfrak{p}\right) }{\mathfrak{p}^{0}d(\mathcal{X})}
\mathcal X\right)  + i   \frac{ 3 t^2
}{2 d(\mathcal X)} \mathcal X \label{Attr-2}
\eeq
which matches   \cite[Formula (24)]{Sh} (\footnote {There is just a sign change that is  due to  our different assumptions. In fact,   according to them, we have  $d(y) > 0$ -- and not $d(y) < 0$ -- at the points $z = x +i y$ of $ \cS$.}).
\end{remark}
The local invertibility property established  in Theorem \ref{main1} has in practice the following meaning.  Assume that   $\gp^I$,  $\gq_{J}$  are  the values of   the  magnetic and electric  charges  of  a BPS  black hole  and that  $Z_o = t_o e^{i \q_o} := Z(r_o)$  and
 $z_o := z(r_o)$ are  the  corresponding values of the central charge and of the scalar fields map  at the horizon $r_o = 0$.  Since  the result  {\it does not}
guarantees  that   $\gf|_{\cA_\pm}$ is    {\it globally} invertible,  it  does not  exclude the  possibility that there is some choice for the cubic polynomial $d$, which allows the existence of  several different     BPS black holes, all of them with  the same    electric and magnetic  charges,  but   also each of them with  distinct horizon values for the  central charge  or  scalar fields. In other words,    for an appropriate choice of $d$, it might be that  there  are  distinct BPS black holes  with charges and horizon values for  the  central charge and the scalar fields   with
$$(p^I, q_K) {=} (p'{}^I, q'_K) {=}  (p''{}^I, q''_K) {=} \ldots \  \text{ but } \
 (Z_o, z_o )  {\neq} (Z_o', z_o' )  {\neq}  (Z_o'', z_o'') \neq \ldots\ .$$
However, even if this is the case,   the pairs $(z_o, Z_o) \neq (z_o', Z'_o) \neq (z_o'', Z''_o) \neq \ldots$  constitute   a  {\it discrete set} of points in $\cA_+ \cup \cA_- \subset \bC^* \times \cS_\cT$. Indeed,  for  each such pair {\it  there must be  a neighbourhood on which the BPS map   is  one-to-one}.
\par
We finally stress the fact that the  proof of   Theorem  \ref{main1}   shows  that there exists  a {\it global} inverse each map map $\gf: \cA_\pm \to  \gf(\cA_{\pm}) \subset \bR^{2n+2}$ if and only if the restriction to $\cV$ of   $h $ admits a global inverse, namely
\begin{cor} \label{thecorollary}  If the  restriction    $h|_{\cV} : \cV \to  h(\cV) \subset \bR^{n}{}' $ admits an  inverse  $(h|_{\cV})^{-1}: h(\cV) \to \cV$,
then each of the two maps $\gf|_{\cA_{\pm}}: \cA_\pm \to   \gf(\cA_{\pm}) \subset \bR^{2n+2}$ is a diffeomorphism onto its  image, with inverse given by  \eqref{5-1} -- \eqref{7-1}.
\end{cor}
\par
\smallskip
  \subsection{The entropy of  BPS black holes in case of  homogeneous  scalar manifolds}
  \label{conclusion}
By Corollary \ref{thecorollary},  Theorem \ref{theo32},  Theorem \ref{thetheorem}, Lemma \ref{lemma29},   if the scalar manifold
 $\cS_\cT = \bR^n + i \cV$, $\cV = \bR_+ {\cdot} \cT \subset  \{ d  > 0\}$,   is  homogeneous and determined by an irreducible  invariant cubic polynomial  $d =  k d_\cV: \bR^n \to \bR$,  then
$\cV$ is a special Vinberg cone and  the corresponding quadratic map  $h|_\cV:  \cV \subset \bR^n \to  \cV' \subset  \bR^{n*}$
 is globally invertible with inverse given by
 \beq \label{poropo} 
 \begin{split}& h^{-1}(y) =  \frac{1}{\sqrt{(d'+\gd')(y)}} (h' + \gh' )(y) \\
 & \qquad \text{with}\ d'  +  \gd' = \text{dual  invariant   cubic rational function } \\
 & \qquad  \text{and} \  \left(h' + \gh' \right)(y)(\cdot)=  \frac{1}{3}\frac{d}{ds} \big((d' + \gd')(y + s (\cdot))\big)\big|_{s = 0} 
  \end{split} \eeq
As a consequence,   the
entropy $S = \pi |Z|^2$ of an extremal BPS black hole with magnetic charge  $p^0 \neq 0$  is uniquely determined by the  black hole  charges $(p^I, q_J) = (p^0, p, q_0, q)$
by means of the formula  \eqref{7-1bisbis}, which provides $t^2 = |Z|^2$. More precisely, since  in the homogeneous case the map $ h^{-1}$ is given by   \eqref{poropo}, 
the term $\langle  \D, \mathcal X\rangle$  in \eqref{7-1bisbis} can be written as
\beq \label{534}
\begin{split} 
&  \langle  \D, \mathcal X\rangle =  3^{\frac{3}{2}} \left \langle h\left( p \right) - \frac{1}{3} p^0q,  h^{-1} \left( h( p ) - \frac{1}{3}p^0q\right)\right\rangle
=  \\
& = 3^{\frac{3}{2}} (1 + \ga)
\sqrt{d' \left( h( p ) - \frac{1}{3}p^0q\right) }  +  3^{\frac{3}{2}} \gb
\\
&  \text{where} \ \ \ga \=  \frac{\sqrt{d' \left( h( p ) - \frac{1}{3}p^0q\right) } }{\sqrt{d'\left( h( p ) - \frac{1}{3}p^0q\right) +  \gd'\left( h( p ) - \frac{1}{3}p^0q\right)}} - 1\ \ \\
& \text{and} \ \  \gb \= \frac{  \bigg \langle h\left( p \right) - \frac{1}{3} p^0q,  
 \gh' \left( h( p ) - \frac{1}{3}p^0q\right)
\bigg\rangle}{\sqrt{d'\left( h( p ) - \frac{1}{3}p^0q\right) +  \gd'\left( h( p ) - \frac{1}{3}p^0q\right)}} 
\end{split}
\eeq
 It follows that
\begin{multline} \label{5.36} S = \pi \sqrt{I_4} \qquad \text{with}\\
   I_4 =  \frac{4 }{ (p^0)^2}  \bigg(   (1 + \ga)
\sqrt{d' \left( h( p ) - \frac{1}{3}p^0q\right) }  + \gb\bigg)^2 - \\
- \frac{1}{(p^0)^2}  \bigg(( q_0 p^0  +  \langle  q, p \rangle)p^0
-  2 d(p)
\bigg)^2
\end{multline}
 It is a remarkable fact that, when $\gd = 0$ and thus $\ga = \gb = 0$ (i.e. when the cone $\cV$ is self-adjoint) this  homogeneous rational function of degree $4$  is actually a   {\it quartic polynomial}. Indeed, it is straightforward to check that, due to
 the identity \eqref{formula},  when $\gd = \ga = \gb  = 0$
 \begin{align*}
 &  \frac{4}{(p^0)^2} d'\left(h( p ) - \frac{1}{3}p^0q\right) =  \frac{4}{(p^0)^2} \langle h'(h(p)), h(p) \rangle  - \\
 & \hskip 5 cm   - \frac{4}{p^0}\langle h'(h(p)), q\rangle + \frac{4}{3}  d'(q)\rangle
  - \frac{4 p^0}{ 3^3} \langle h'(q), q\rangle = \\
 &\hskip 3 cm  = \frac{4}{(p^0)^2} (d(p))^2  - \frac{4}{p^0} d(p) \langle p, q\rangle  + \frac{4}{3}  \langle h(p), h'(q)\rangle -   \frac{4 p^0}{3^3}d'(q)\ ,\\
  & -  \frac{1}{(p^0)^2} \bigg( (p^0(p^0 q_0 +  \langle  q, p \rangle)^2 -  2  d(p)\bigg)^2  = \\
  & \hskip 2.5 cm =  - ( q_0 p^0  +  \langle  q, p \rangle)^2 + 4   q_0  d(p)   + \frac{4}{p^0} \langle  q, p \rangle d(p) -
  \frac{4}{(p^0)^2} (d(p))^2\ ,
   \end{align*}
 which imply
\beq \label{theI4}
   I_4 =
    -  ( q_0 p^0  +  \langle  q, p \rangle)^2 + 4  q_0  d(p)  -     \frac{4}{27} p^0   d'(q) + \frac{4}{3}  \langle h(p),     h'(q)\rangle\ .
\eeq
Summing up, we have the following final result:
\begin{theo}  In  ungauged  $N=2$  $D=4$ supergravity  with homogeneous scalar manifold $\cS$ (not necessarily symmetric)  given by an  irreducible cubic polynomial $d$ (see \S \ref{section31}), the entropy of  the extremal BPS black holes  defined in  \S \ref{section3.3} is expressed in terms of their magnetic and electric charges $(p^0, p, q_0, q)$ by
\beq \label{ecco} \begin{split}
& S = \pi \sqrt{I_4}\ \text{with}\ I_4 \ \text{as in \eqref{5.36} (reducing to \eqref{theI4} in case $\cS$ is symmetric)}\ .
\end{split}
\eeq
\end{theo}
\par
\medskip
 Note  that, in the cases
in which $\cS$ is  symmetric, \eqref{ecco} coincides with  the entropy formula  which was so far  known \cite{FG}.  To check this, one should first  recall that, in  the physics literature, the prepotential  is usually assumed to  have the form
$F(X) = \frac{1}{3!} \frac{D_{abc} X^a X^b X^c}{X^0}$, meaning that $D$ and our cubic polynomial $d$  are related  by $  d= \frac{1}{6} D $.   Then, we remind that
 {\it if the scalar manifold $\cS = \bR^n + i \cV$ is symmetric}, it has for long time  known that there exists a  symmetric contravariant cubic tensor $D'(Y^\flat) = D'{}^{abc} Y_a Y_b Y_c$ satisfying the  ``adjoint identity" \cite{GST, CV}
\beq D'(D(X), D(X)) = \frac{4}{3} D(X) X\ .\eeq
By substituting $D = 6  d =  6 d^*$ and comparing with \eqref{formula}, we see that  the contravariant tensor $D'$ used in physics literature  is related with our invariant dual cubic polynomial $d'$  by
$$d' = \frac{9}{2} D'\ .$$
 Plugging all this  in \eqref{theI4},  the expression  reduces to
\beq \label{539}
   I_4 =
    -  ( q_0 p^0  +  \langle  q, p \rangle)^2 + 4 q_0  I_3(p)  -    4 p^0  I_3(q) +   4 \{I_3(p), I_3(q)\}\  ,\eeq
where
\beq \nonumber
\begin{split}
&I_3(p) \= \frac{1}{3!} D(p)\ ,\quad   I_3(q) \= \frac{1}{3!} D'(q)\ ,\\
   & \text{and} \ \{\cdot ,  \cdot \} \ \text{stands for the standard Poisson bracket in}\ \bR^n \times \bR^{n*}\ ,\\
 & \text{which gives }\ \  \  \{I_3(p), I_3(q)\} =  \frac{1}{4}  \langle D(p, p, \cdot),     D'(q, q, \cdot)\rangle = \frac{1}{3} \langle h(p), h'(q)\rangle\ .
 \end{split}
\eeq
The \eqref{539} is precisely  the formula  so far  known only for the cases with   symmetric scalar manifolds.

\vskip 0.5truecm
\hbox{\parindent=0pt\parskip=0pt
\vbox{\baselineskip 9.5 pt \hsize=3.7truein
\obeylines
{\smallsmc
Dmitri V.  Alekseevsky
Institute for Information Transmission Problems
B. Karetnuj per. 19
Moscow 127051, Russia
\&
    University of Hradec Králové
    Faculty of Science
    Rokitanského 62
   500 03 Hradec Králové, Czech Republic
}\medskip
{\smallit E-mail}\/: {\smalltt dalekseevsky@iitp.ru}
}
}
\vskip 1 truecm
\hbox{\parindent=0pt\parskip=0pt
\vbox{\baselineskip 9.5 pt \hsize=3.1truein
\obeylines
{\smallsmc
Alessio Marrani
Instituto de Física Teorica, 
Departamento de Física,
Universidad de Murcia, Campus de Espinardo, 
E-30100, Spain
 }\medskip
{\smallit E-mail}\/: {\smalltt alessio.marrani@um.es}}

}
\vskip 1 truecm
\hbox{\parindent=0pt\parskip=0pt
\vbox{\baselineskip 9.5 pt \hsize=3.1truein
\obeylines
{\smallsmc
Andrea Spiro
Scuola di Scienze e Tecnologie
Universit\`a di Camerino
Via Madonna delle Carceri
I-62032 Camerino (Macerata)
Italy
}\medskip
{\smallit E-mail}\/: {\smalltt andrea.spiro@unicam.it
}
}
}

\end{document}